	\documentclass[pra,showpacs,twocolumn,floatfix,aps]{revtex4}
	\usepackage{amsmath}
	\usepackage{makeidx}
	\usepackage{amssymb}
	\usepackage{graphicx}

	\usepackage[usenames]{color}
	\usepackage[normalem]{ulem}

	\usepackage{hyperref}
	\usepackage[T1]{fontenc} 

\begin{document}


	\title{Three-dimensional vortex-bright solitons in a spin-orbit coupled spin-$1$ condensate}

	\author{Sandeep Gautam\footnote{sandeep@iitrpr.com}}
        \affiliation{Department of Physics, Indian Institute of Technology Ropar, Rupnagar, Punjab 140001,
                     India
                     }
	\author{S. K. Adhikari\footnote{adhikari44@yahoo.com, 
		URL  http://www.ift.unesp.br/users/adhikari}}
	\affiliation{Instituto de F\'{\i}sica Te\'orica, Universidade Estadual
		     Paulista - UNESP,  01.140-070 S\~ao Paulo, S\~ao Paulo, Brazil}
	      

	\date{\today}
	\begin{abstract}

	We demonstrate stable and metastable vortex-bright solitons in a 
        three-dimensional spin-orbit-coupled three-component hyperfine spin-1  Bose-Einstein 
        condensate (BEC) using numerical solution and variational approximation of a 
        mean-field model. The spin-orbit coupling provides attraction  to 
        form vortex-bright solitons in both attractive and repulsive  spinor BECs.
        The ground state of these vortex-bright solitons is axially  symmetric for  weak polar 
	interaction. For a sufficiently strong ferromagnetic interaction,
	we observe the emergence of a fully asymmetric vortex-bright soliton as the ground 
	state. We also numerically investigate  moving solitons. The present 
	mean-field model is not Galilean invariant, and 
	we use a Galilean-transformed  mean-field model 
	for generating the moving solitons.

	\end{abstract}

	\maketitle


	\section{Introduction}
	\label{Sec-I}
	 A bright soliton, which arises due to a cancellation of the effects produced 
	by non-linear and dispersive terms in the Hamiltonian, is a self-reinforcing 
	solitary wave which maintains its shape while moving at a constant speed. 
	Studies on the solitons have been done in a broad array of systems which include,
	among others, water waves, non-linear optics \cite{Kivshar}, ultracold quantum gases 
	including spinor Bose-Einstein condensates (BECs) \cite{Inouye,li, rb, Perez-Garcia,Ieda}. 

	The spin-orbit (SO) coupling, the coupling between the spin and the center of mass motion of
	the atoms, is absent in the neutral atoms \cite{stringari}. Nevertheless,
	a suitable modification of the atom-light interaction can generate a non-Abelian
	gauge potential \cite{Dalibard}, thus subjecting the neutral atoms to the SO coupling.
	Guided by this idea, Lin {\em et al.} \cite{Lin} experimentally generated an SO coupling with 
	equal strengths of Rashba \cite{Rashba} and Dresselhaus \cite{Dresselhaus} terms 
	in a BEC of $^{87}$Rb in the two-component pseudo-spin-1/2 configuration, where one of the 
        three spin components of the hyperfine spin-1 state of $^{87}$Rb was removed from the experiment. 
        This was achieved by dressing two of $^{87}$Rb spin states 
	from within its ground electronic manifold ($5S_{1/2}, F = 1$) with a 
	pair of lasers \cite{Lin}. More recently, SO coupling has been realized experimentally by Campbell {\em et al.} \cite{Campbell}
	 with the three hyperfine spin components 
	of $^{87}$Rb atoms.  A lot of experimental studies
	have been done on SO-coupled BECs in recent years \cite{Aidelsburger}.

	It has been shown theoretically  that the SO-coupled quasi-one-dimensional
	(quasi-1D) \cite{Salasnich,rela}, quasi-two-dimensional (quasi-2D) \cite{Xu,Sakaguchi}, 
        and three-dimensional (3D) \cite{pu} pseudo-spin-1/2 BECs  can support
	solitonic structures. Bright solitons in   SO-coupled
	three-component	quasi-1D spin-1 \cite{Liu,Gautam-3} and five-component
	spin-2 BECs  \cite{Gautam-4} have also been
	theoretically investigated in addition to those in the three-component
	quasi-2D spin-1 BEC \cite{Gautam-5}.

In this paper, we demonstrate stable and metastable stationary and moving 3D 
vortex-bright solitons in a three-component  SO-coupled hyperfine spin-1 BEC 
using a variational approximation and numerical solution of the mean-field 
Gross-Pitaevskii (GP) equation \cite{Ohmi}. The effect of SO coupling on both 
an attractive and a weakly repulsive spinor BEC is to introduce attraction so 
as to form a soliton \cite{Sakaguchi}. We find  metastable vortex-bright 
solitons for $a_0+2a_2<0$, where $a_0$ and $a_ 2$ are the $s$-wave scattering 
lengths in the total spin 0 and 2 channels. The solitons can be stable for 
$a_0+2a_2\ge 0$. In the former case, the spinor BEC without SO-coupling is 
attractive, and the collapse cannot be stopped unconditionally thus producing 
only  metastable solitons in the SO-coupled BEC for the number of atoms 
smaller than a critical number as in the case of a single-component 
quasi-one-dimensional attractive BEC \cite{li,pg}.  In the latter case the 
spinor BEC can be purely repulsive and,  because of this repulsion, it is 
possible to stop the collapse to form a stable soliton. In general, the 
implementation of SO-coupling in the three-component spin-1 BEC is more 
complicated than the same in the two-component pseudo-spin-1/2 BEC  from both 
theoretical \cite{gtm} and experimental \cite{Campbell} points of view.
Thus the present study goes beyond a previous investigation of 3D metastable 
bright solitons in pseudo-spin-1/2  BEC \cite{pu}. Moreover, the parameter 
domain $(a_0+2a_2\ge 0)$ that leads to stable 3D vortex-bright solitons in a 
SO-coupled spin-1 BEC was not considered before in this context.  	

        We observe that for small strengths of SO coupling, which we use in 
this investigation, the ground state vortex-bright soliton in the polar domain
$(a_2>a_0)$ has an antivortex and a vortex in $m_f= +1$ and $m_f = -1$ 
components, respectively, and a Gaussian-type structure in the $m_f=0$ 
component. The phase singularities in the $m_f= \pm 1$ components always 
coincide leading to axisymmetric density profiles for the component 
wavefunctions. We use phase-winding numbers \cite{Gautam-5} (angular momenta 
per particle in an axisymmetric system) of the three component wavefunctions 
to denote a vortex or an antivortex \cite{Mizushima}. In terms of 
phase-winding numbers associated with the spin components $m_f = +1,0,-1$ 
\cite{Mizushima}, this ground state vortex-bright soliton in the polar domain 
can be termed as symmetric and denoted $(-1,0,+1)$ corresponding to an 
anti-vortex in component $m_f = +1$ and a vortex in component $m_f = -1$. 
In the ferromagnetic domain ($a_0>a_2$), in addition to the axisymmetric 
vortex-bright solitons, asymmetric vortex-bright solitons with non-coinciding 
phase singularities in $m_f= \pm 1$ can also emerge as the ground state below 
a critical value of spin-exchange interaction parameter. 
In addition to this, we have also identified  stationary excited axisymmetric
vortex-bright solitons of type $(0,+ 1,+ 2)$ in both polar and ferromagnetic 
domains. Besides stationary vortex-bright solitons, we have also investigated 
the dynamically	stable moving vortex-bright solitons of the SO-coupled spin-1 
BEC using the Galelian-transformed coupled GP equations \cite{rela,Sakaguchi,Liu,Gautam-3}.

	The paper is organized as follows. In Sec. \ref{Sec-IIA}, we describe 
the mean-field coupled Gross-Pitaevskii (GP) equations with Rashba SO coupling
used to study the vortex-bright solitons in a spin-1 BEC. This is followed by 
a variational analysis of the stationary axisymmetric $(-1,0,+1)$ 
vortex-bright solitons in Sec. \ref{Sec-IIB}. In Sec. \ref{Sec-III}, we 
provide the details of the numerical method used to solve the coupled GP 
equations with SO coupling. We discuss the numerical results for axisymmetric 
vortex-bright solitons in Sec. \ref{Sec-IVA}, asymmetric solitons in 
Sec. \ref{Sec-IVB}, and moving solitons in Sec. \ref{Sec-IVD}. Finally, 
in Sec. V, we give a summary of our findings.


\section{Spin-Orbit coupled   BEC vortex-bright soliton}
\label{Sec-II} 
\subsection{Mean-field equations}
\label{Sec-IIA}
For the study of a 3D vortex-bright soliton, we consider a trapless spin-1 
spinor BEC. The single particle Hamiltonian of the BEC with Rashba 
\cite{Rashba} SO coupling is \cite{H_zhai}
\begin{equation}
H_0 = \frac{p_x^2+p_y^2+p_z^2}{2m}  + \gamma p_x \Sigma_x +
      \gamma p_y \Sigma_y+\gamma p_z\Sigma_z, \label{sph} 
\end{equation}
where $p_x = -i\hbar\partial/\partial x$, $p_y = -i\hbar\partial/\partial y$, 
and $p_z = -i\hbar\partial/\partial z$ are the momentum operators along 
$x$, $y$, and $z$ axes, respectively, $m$ is the mass of each atom and 
$\Sigma_x$, $\Sigma_y$, and $\Sigma_z$ are the irreducible representations of 
the $x$, $y$, and $z$ components of the spin matrix, respectively,
\begin{align}
\Sigma_x&=\frac{1}{\sqrt 2} \begin{pmatrix}
          0 & 1 & 0 \\
	  1 & 0  & 1\\
	  0 & 1 & 0
	\end{pmatrix}, \quad  
\Sigma_y=\frac{1}{\sqrt 2 i} \begin{pmatrix}
        0 & 1 & 0 \\
	-1 & 0  & 1\\
	0 & -1 & 0
	\end{pmatrix},\nonumber\\
	&   
\Sigma_z=\begin{pmatrix}
	1 & 0 & 0 \\
	0 & 0  & 0\\
	0 & 0 & -1
	\end{pmatrix},\label{spin_matrices}
\end{align}
and $\gamma$ is the strength of SO coupling. In the mean-field approximation,
the SO-coupled 3D spin-1 BEC of $N$ atoms is described by the following set 
of three coupled GP equations, written here in dimensionless form, for 
different spin components $m_f=\pm 1,0$ \cite{Ohmi,Kawaguchi}
 \begin{align}
i\frac{\partial \psi_{\pm 1}(\mathbf r)}{\partial t} &=
	 {\cal H}\psi_{\pm 1}(\mathbf r) 
	\pm   c^{}_1F_z\psi_{\pm 1}(\mathbf r) 
	+ \frac{c^{}_1}{\sqrt{2}} F_{\mp}\psi_0(\mathbf r)\nonumber\\
	&-\frac{i\gamma}{\sqrt{2}}\left(
  \frac{\partial\psi_0}{\partial x}\mp i\frac{\partial\psi_0}{\partial y}
  \pm\sqrt{2}\frac{\partial\psi_{\pm 1}}{\partial z}\right),
	 \label{gps-1}\\
	i \frac{\partial \psi_0(\mathbf r)}{\partial t} &=
	{\cal H}\psi_0(\mathbf r)  
	+ \frac{c_1}{\sqrt 2} [F_{-}\psi_{-1}(\mathbf r) 
	+F_{+}\psi_{+1}(\mathbf r)]\nonumber\\
	&-\frac{i \gamma}{\sqrt{2}}\Bigg(\frac{\partial\psi_{1}}{\partial x} 
        +i \frac{\partial\psi_{1}}{\partial y} 
	  +\frac{\partial\psi_{-1}}{\partial x}-
         i\frac{\partial\psi_{-1}}{\partial y}\Bigg)
	\label{gps-2}, 
\end{align}
where ${\bf r}\equiv \{x,y,z\}$, ${\bf F}\equiv\{F_x,F_y,F_z\}$ is a vector 
whose three components are the expectation values of the three spin-operators 
over the multicomponent wavefunction, and is called the spin-expectation 
value \cite{Kawaguchi}. Also,
	\begin{align}&
	F_{\pm}\equiv  F_x \pm i F_y=
	\sqrt 2[\psi_{\pm 1}^*(\mathbf r)\psi_0(\mathbf r)
	+\psi_0^*(\mathbf r)\psi_{\mp 1}(\mathbf r)]\label{fpmspin1}, \\
	&  F_z= n_{+1}(\mathbf r)-n_{-1}(\mathbf r)\label{fzspin1},
	\quad
	{\cal H}= -\frac{\nabla^2}{2} +c_0 n(\bf r),\\
	&c_0 = \frac{4 N \pi (a_0+2 a_2)}{3 l_0},~
	c_1 = \frac{4N \pi ({a_2-a_0})}{3 l_0},\label{nonlin} \\
&
\nabla^2=\frac{\partial ^2}{\partial x^2} + 
      \frac{\partial ^2}{\partial y^2} + \frac{\partial ^2}{\partial z^2}, \label{nabla_q2d}
\end{align}
where $n_j({\bf r})=|\psi_j({\mathbf r})|^2$ with $j=\pm 1, 0$ are the 
component densities, $n({\bf r})=\sum_{j}n_j({\bf r})$ is the total density,
and asterisk denotes complex conjugate. The normalization condition satisfied 
by the component wavefunctions $\psi_j$ is 
\begin{equation}\label{norm}
  \int \sum_j n_j({\bf r}) d{\bf r} =1.
\end{equation}   
All quantities in Eqs. (\ref{gps-1})-(\ref{nabla_q2d}) are dimensionless.
This is achieved by writing length, density, time, and energy in units of
$l_0$, $l_0^{-3}$, $m l_0^2/\hbar$, and  $\hbar^2 /(ml_0^2)$, respectively,
where $l_0$ is a scaling length and can be taken as $l_0=1$ $\mu$m. 
The energy of an atom in dimensionless unit is
\begin{widetext}
\begin{align} 
E &=\int d {\bf r} \biggr[\frac{1}{2}\biggr\{\sum_{j=-1}^1 \left|\nabla 
\psi_j\right|^2 +  (c_0 n^2+c_1 |\mathbf F|^2)\biggr\}-	
\frac{i\gamma}{\sqrt{2}}\psi_0^*
 \left(\frac{\partial \psi_{+1}}{\partial x} + 
\frac{\partial \psi_{-1}}{\partial x}\right)
 + \frac{\gamma}{\sqrt{2}}\psi_0^*\left(\frac{\partial \psi_{+1}}{\partial y} 
- \frac{\partial \psi_{-1}}{\partial x}\right)\nonumber  \\ 
& -\frac{i\gamma}{\sqrt{2}} \left(\psi_{+1}^*+\psi_{-1}^*\right)    
\frac{\partial \psi_0}{\partial x} -\frac{\gamma}{\sqrt{2}} \left(\psi_{+1}^*-
\psi_{-1}^*\right)\frac{\partial \psi_0}{\partial y} - i\gamma 
\left( \psi_{+1}^*\frac{\partial \psi_{1}}{\partial z} - 
\psi_{-1}^*\frac{\partial \psi_{-1}}{\partial z}\right)\biggr].
 \label{energy}
\end{align}
\end{widetext}

It is instructive to analyze the SO-coupled system in the absence
of interactions, i.e. $c_0=c_1= 0$. Then, using Eqs. (\ref{sph})-(\ref{spin_matrices}), 
the single particle SO-coupled Hamiltonian of the system is
\begin{align}
H_0 &= 
\begin{pmatrix} 
 -\frac{\nabla^2}{2} -i \gamma \frac{\partial}{\partial z}&   
-\frac{i\gamma}{\sqrt{2}} \partial_{-}&0 \\
-\frac{i\gamma}{\sqrt{2}} \partial_{+} & -\frac{\nabla^2}{2} & -\frac{i\gamma}{\sqrt{2}}\partial_{-}\\
0 & -\frac{i\gamma}{\sqrt{2}}\partial_{+} & -\frac{\nabla^2}{2} + i \gamma\frac{\partial}{\partial z}
\end{pmatrix},
\end{align}
where $ \partial_{\pm}= \left(\frac{\partial}{\partial x} \pm i \frac{\partial}{\partial y}\right)$.
The minimum eigen energy of the single particle Hamiltonian $H_0$ is $-\gamma^2/2$ and
corresponds to $|{\bf k}|\equiv \sqrt{k_x^2+k_y^2+k_z^2} = \gamma$, and the
(unnormalized) eigen function corresponding to this energy is
\begin{align}
\Psi & \sim \frac{1}{2}
\begin{pmatrix} 
 e^{-i\varphi}(|{\bf k}|-k_z)^2/k_{\rho}^2\\
 -\sqrt{2}(|{\bf k}|-k_z)/k_{\rho}\\
e^{i\varphi}
\end{pmatrix}
e^{i (k_x x+k_yy+k_zz)},
\end{align}
where $\varphi=\tan^{-1}(k_y/k_x)$ is
the angle made by the projection of $\bf k$ on the $xy$ plane with the $x$ axis and $k_{\rho} = \sqrt{k_x^2+k_y^2}$. 
A general circularly symmetric solution can be obtained by considering the superposition of degenerate eigen functions with
fixed $|k_z|$ and all possible values of $\varphi$, i.e.
\begin{widetext}
\begin{align}
\,  \Psi_{\rm gen} &\sim \sum_{k_z = |k_z|,-|k_z|}\int_0^{2\pi} \frac{1}{2}
\begin{pmatrix} 
 e^{-i\varphi}(|{\bf k}|-k_z)^2/k_{\rho}^2\\
 -\sqrt{2}(|{\bf k}|-k_z)/k_{\rho}\\
e^{i\varphi}
\end{pmatrix}
 e^{i (k_{\rho} x \cos\varphi + k_{\rho} y \sin\varphi  +k_zz)} d\varphi,\end{align}
\end{widetext}
\begin{align}
                \sim \sum_{k_z = |k_z|,-|k_z|}\begin{pmatrix}
    \pi  e^{i (k_z z+\theta) } J_1(k_{\rho} \rho) (|{\bf k}|-k_z)^2/k_{\rho}^2\\
    \pi  e^{i k_z z } J_0(k_{\rho} \rho)(|{\bf k}|-k_z)/k_{\rho}\\
   -\pi  e^{i (k_z z-\theta) } J_1(k_{\rho } \rho)
\end{pmatrix},\label{psi_gen}
\end{align}
where $\theta = \tan^{-1}({y/x})$, $\rho = \sqrt{x^2+y^2}$, and
$J_{0}(k_\rho \rho)$ and $J_{1}(k_\rho \rho)$ are the Bessel functions of first kind
of order 0 and 1, respectively. In the asymptotic region, $\rho\rightarrow\infty$, $J_{0}(k_\rho \rho) \sim \sqrt{2/(\pi k_\rho \rho)}\cos(k_\rho \rho-\pi/4)$
and $J_{1}(k_\rho \rho) \sim \sqrt{2/(\pi k_\rho \rho)}\sin(k_\rho \rho-\pi/4)$ demonstrating the oscillatory 
nature of the wave function. 
 The actual values of $k_\rho$ and $k_z$ will depend upon the full
minimization of energy functional Eq. (\ref{energy}) satisfying $k_\rho^2+k_z^2 = \gamma^2$.

\subsection{Vortex-bright  soliton}
\label{Sec-IIB}
We demonstrate  the existence of two types of metastable and stable 
low-energy stationary axisymmetric vortex-bright solitons, classified using 
phase-winding numbers as $(-1,0,+1)$ and $(0,+1,+2)$, and an asymmetric 
vortex-bright soliton. Out of the former two, the  $(-1,0,+1)$ vortex-bright
soliton has the lower energy. We find that, in the polar domain ($c_1>0$), 
the $(-1,0,+1)$ vortex-bright soliton  has coinciding phase singularities in
$m_f = \pm 1$ components, which results in axially-symmetric density profiles
of the component wavefunctions and is the ground state. In the ferromagnetic 
domain ($c_1<0$), below a critical $c_1$, in addition to  $(-1,0,+1)$ and 
$(0,+1,+2)$ vortex-bright solitons, we observe the emergence of ground state 
vortex-bright solitons in which an antivortex in the $m_f = +1$ component 
does not coincide with a vortex in the $m_f = -1$ component. This results in 
an asymmetric density profile for the component wavefunctions. The higher 
energy $(0,+1,+2)$ vortex-bright solitons obtained numerically are always 
axially symmetric due to coinciding phase singularities. 
Equations (\ref{gps-1})-(\ref{gps-2}) are invariant under transformations:
$y \rightarrow -y$, $z\rightarrow -z$, $\psi_{m_f}(x,y,z)\rightarrow 
\psi_{-m_f}(x,-y,-z)$. Under these transformations, a symmetric 
$(-1,0,+1)$ vortex-bright soliton transforms to itself, whereas a 
symmetric $(0,+1,+2)$ vortex-bright soliton transforms to $(-2,-1,0)$
vortex-bright soliton. This implies that associated with the $(0,+1,+2)$  
vortex-bright soliton there is always a  degenerate $(-2,-1,0)$  
vortex-bright soliton.

Numerically, we find that the longitudinal magnetization 
${\cal M} = \int[ {n_{+1}({\bf r})-n_{-1}({\bf r})}]d\bf r$ is zero for 
the  symmetric $(- 1,0,+ 1)$  and asymmetric vortex-bright 
solitons, whereas it is, in general, non-zero for the $(0,+ 1,+ 2)$ solitons. 
The $(-1,0,+1)$ vortex-bright soliton  with zero magnetization $\cal M$ 
can be analyzed using the following variational {\em ansatz} \cite{pu}
\begin{align}
\psi_{\pm 1} &=\left(A_1\sigma_z \mp i\sqrt{2} B z \right)  
\frac{\rho e^{-\frac{\rho^2}{2 \sigma_{r}^2} -\frac{z^2}{2 \sigma_{z}^2} 
\mp i \tan^{-1}\frac{y}{x}}}{\pi^{3/4}\sigma_r^2\sigma_z^{3/2}},\label{ansatz1}
\\
\psi_0&=i \frac{A_2}{\pi^{3/4} \chi_r\sqrt{\chi_z}} 
e^{-\frac{\rho^2}{2 \chi_r^2}-\frac{z^2}{2 \chi_z^2}},\label{ansatz2}
\end{align}
where 
$A_1,A_2, B, $ are the variational amplitudes and 
$\sigma_r,\sigma_z,\chi_r$, and $\chi_z$ are the variational widths of the 
Gaussian {\em ansatz}. Normalization condition (\ref{norm}) imposes the 
following constraint 
\begin{equation}           
\label{cons}    2A_1^2+2B^2+A_2^2 =1
\end{equation}
on the variational parameters $A_ 1,A_2$ and $B$.
The  variational energy of the soliton, obtained by substituting 
Eqs. (\ref{ansatz1}) and (\ref{ansatz2}) in Eq. (\ref{energy}), is
 \begin{widetext}
\begin{align}
E  &=    \frac{2 \left(A_1^2+B^2\right)}{\sigma_r^2}
+\frac{ A_1^2+3 B^2}{2\sigma_z^2}+\frac{A_2^2}{4} 
\left(\frac{2}{\chi_r^2}+\frac{1}{\chi_z^2}\right) 
 + \frac{c_0}{\pi^{3/2}}\biggr[ \frac{   A_2^4} {4\sqrt{2} \chi_z \chi_r^2}+
\frac{\left(4 A_1^4+4 A_1^2 B^2+3 B^4\right)}{8\sqrt{2} \sigma_r^2 \sigma_z}  
\nonumber  \\
&+\frac{  {2A_2^2\chi_r^2\left\{B^2 \chi_z^2+A_1^2 \left(\sigma_z^2+
\chi_z^2\right)\right\}}}{\left(\sigma_z^2+\chi_z^2\right)^{3/2}
 \left(\sigma_r^2+\chi_r^2\right)^2}\biggr]-\frac{16 A_1 A_2 \gamma
\sigma_r^2  \chi_r \sqrt{\sigma_z\chi_z}}{\left(\sigma_r^2+\chi_r^2\right)^2 
\sqrt{ \left(\sigma_z^2+\chi_z^2\right)}}-2\sqrt 2 \frac{A_1 B\gamma}
{\sigma_z},\label{E1}
\end{align} 
\end{widetext}
 
This energy is independent of $c_1$ as ${\bf F}({\bf r}) = 0$ for symmetric 
$(-1,0,+1)$ vortex-bright soliton which is consistent with the choice of 
variational {\em ansatz} in Eqs. (\ref{ansatz1})-(\ref{ansatz2}). Energy  
(\ref{E1}) can be minimized with respect to all variational parameters 
subject to the constraint  (\ref{cons})  to obtain the minimum-energy 
ground state.

\begin{figure}[t]
\begin{center}
\includegraphics[trim = 0cm 2cm 0cm 0cm, clip,width=\linewidth,clip]{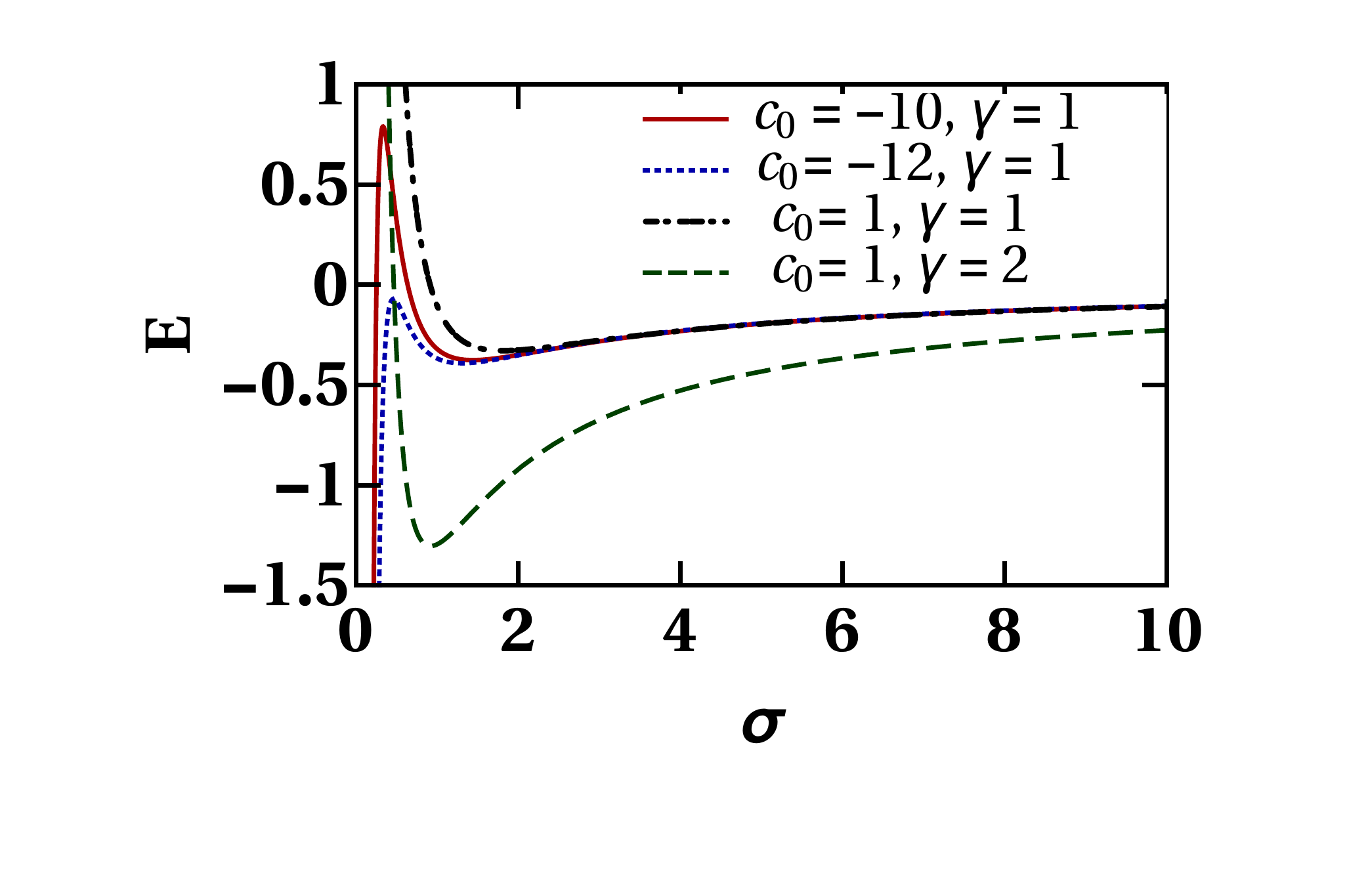}
\caption{ The variational energy (\ref{Eint}) of a self-trapped 3D 
vortex-bright soliton as a function of  width $\sigma$ for the parameters 
$c_0 =1, -10, -12, c _1 > 0, \gamma = 1,2$. A metastable vortex-bright 
soliton corresponds to a local minimum in energy  as appears for $c_0<0$. 
A stable vortex-bright soliton can appear for $c_0\ge 0$. }
\label{Fig1}
\end{center}
\end{figure}

To understand the role of SO coupling in the creation of self-trapped 
vortex-bright solitons, let us assume that the widths of the component 
wavefunction are of the same order of magnitude, say 
$\sigma_r=\sigma_z=\chi_r=\chi_z\approx \sigma$. 
In this case,  energy (\ref{E1}) becomes
\begin{align}
E &= \left[\frac{\left(10 A_1^2+3 A_2^2+14 B^2\right) }{4 \sigma ^2} \right]
  -\left(\frac{2\sqrt{2} A_1 \left( A_2+B\right)  \gamma }{\sigma }\right)
\nonumber\\
  & + \left\{{c_0}\frac{4 A_1^4+2 A_2^4+2 A_2^2 B^2+3 B^4+4 A_1^2 
\left( A_2^2+B^2\right) }{8 \sqrt{2}\pi^{3/2} \sigma ^3}\right\},
\nonumber      
\end{align}
\begin{align}
&\equiv\left[ \frac{C_{1}}{\sigma ^2}\right] -\left(\frac{\gamma C_{2}}
{\sigma}\right) + \left\{\frac{c_0C_{3}}{\sigma ^3}  \right\}
,\label{Eint}
\end{align}
where $C_{1}, C_{2}, C_{3}$ are the functions of $A_1$ and $B$ (as $A_2$
itself is a function of $A_1$ and $B$) and are all greater than zero. If 
we let, $A_1$ and $B$ to assume all possible real values greater than zero, 
then the total energy has a local minimum at
\begin{equation}
\sigma = \frac{C_1+\sqrt{C_1^2+3\gamma c_0C_2C_3}}{\gamma C_3},
\end{equation}
provided $\gamma>0$ and $C_1^2 +3\gamma c_0 C_2 C_3>0$, the
latter inequality implies that the dispersive effects are strong enough 
to prevent the collapse of the system. 

To see explicitly, how a metastable (stable) self-trapped symmetric 
$(-1,0,+1)$ vortex-bright soliton corresponds to local (global) minimum of 
energy, we minimize energy (\ref{E1}) and calculate all seven variational 
parameters: $\sigma_r, \sigma_z, \chi_r, \chi_z, A_1,A_2  $ and $B$. Using the values of $A_1, B$, and 
$A_2$ so obtained, we calculate the variation of energy (\ref{Eint}) as a 
function of the width $\sigma$. The resulting $E$ versus $\sigma $ curves are 
shown in Fig. \ref{Fig1} for different $c_0$ and a fixed SO coupling 
$\gamma$ illustrating a local minimum and also  the collapse as 
$\sigma \rightarrow 0$ for $c_0<0$: energy $E\to -\infty$ as 
$\sigma \rightarrow 0$. For $c_0\ge 0$, there could be a global minimum with 
no possibility of collapse and one has a stable soliton. We see 
in Fig.  \ref{Fig1} that the minimum of the  energy is more pronounced for a larger 
spin-orbit coupling $\gamma$, thus leading to stronger binding.
The case $c_0\ge 0, 
c_1=\gamma=0$ corresponds to a repulsive spinor condensate. For 
$0>c_0> c_{\mathrm{crit}}$,  only a local minimum of energy is possible and 
the vortex-bright soliton is  metastable, whereas, for 
$c_0< c_{\mathrm{crit}}$, no localized soliton is possible and the system 
collapses. In contrast,	a quasi-two-dimensional (quasi-2D) self-trapped 
vortex-bright soliton is always stable and corresponds to a global minimum of 
energy \cite{Gautam-5}.
 
\begin{figure}[t]
\begin{center}
\includegraphics[trim = 2mm 0cm 0cm 0cm, clip,width=0.49\linewidth,clip]{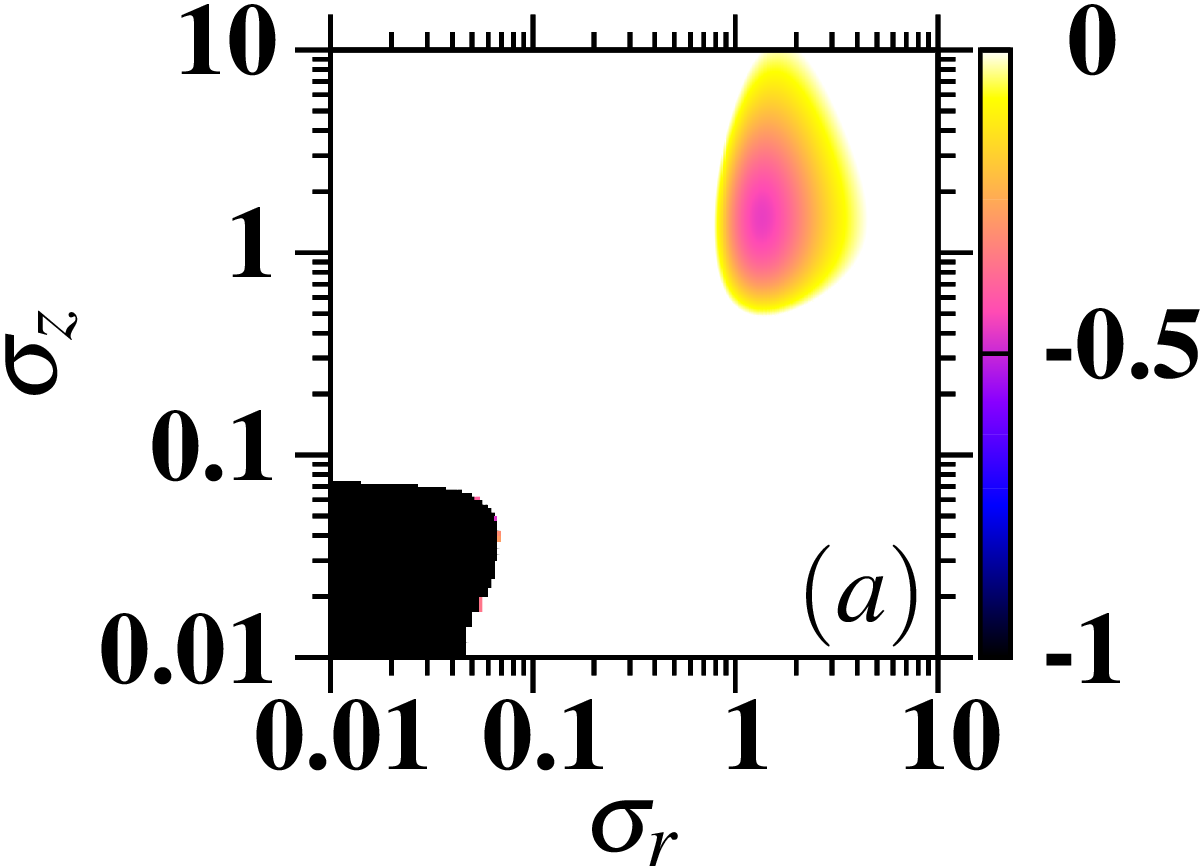}
\includegraphics[trim = 0mm 0cm 0cm 3mm, clip,width=0.49\linewidth,clip]{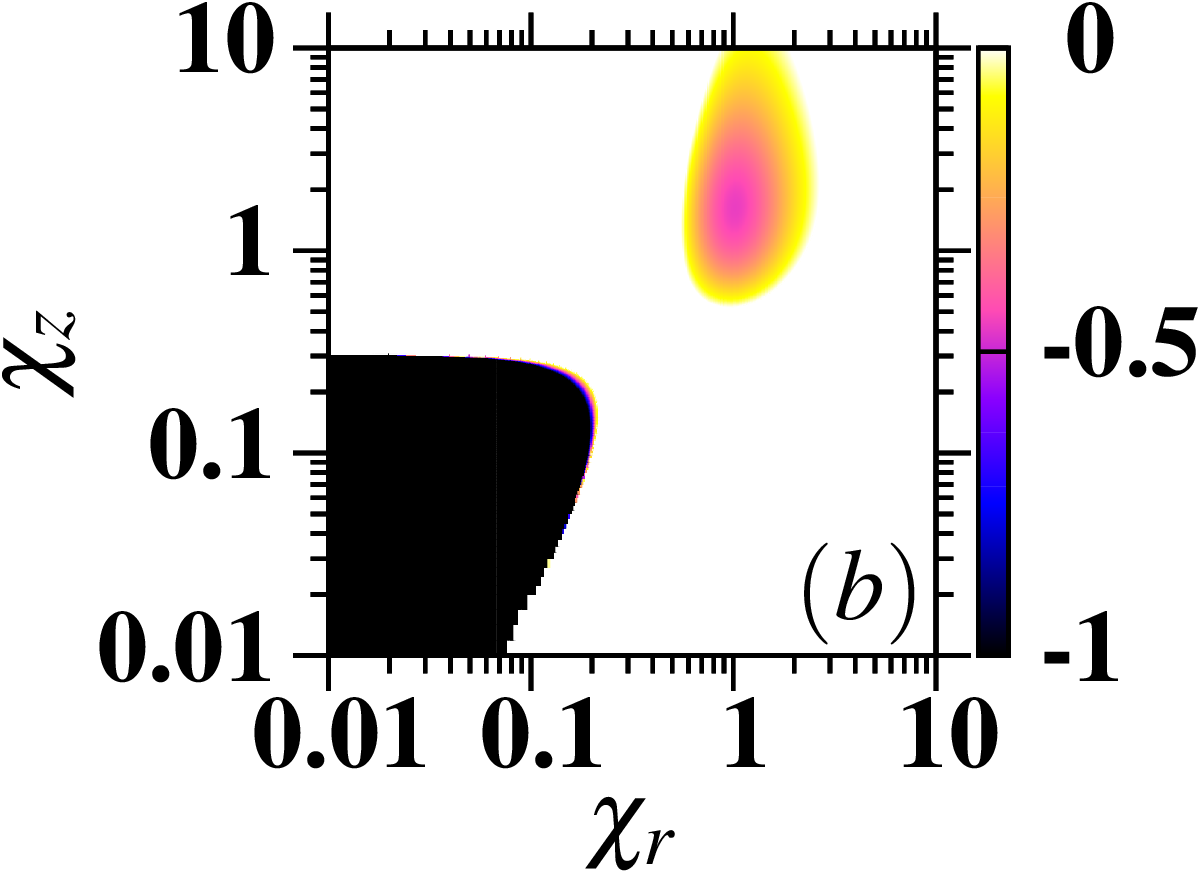}
\caption{(a) Contour plot variational energy as a function of radial and 
axial widths of $m_f = \pm 1$ components, (b) the same as a function of 
radial and axial widths of $m_f = 0$ component. The local minimum of energy 
corresponding to the vortex-bright solitons as well as the collapse near the 
origin are highlighted. The interaction parameters are $c_0 = -10, c _1 > 0, 
\gamma = 1$. See the text for further details.}
\label{Fig2}
\end{center}
\end{figure}

One can also look at the variation of energy  (\ref{E1})  by fixing $A_1,B$, 
and hence $A_2$ and two variational parameters characterizing the widths of 
the components as a function of the remaining two variational widths. For this
purpose, we find the variational parameters $A_1, B, A_2,\sigma_r, \chi_r,
\chi_z$ and $\sigma_z$ corresponding to the minimum of energy  (\ref{E1}), 
which  is a function of these parameters. Fixing the parameters 
$A_1, B, A_2, \chi_r,$ and $\chi_z$ at the values corresponding the minimum 
of energy (\ref{E1}), we consider energy as a function of the widths 
$\sigma_r$ and $\sigma_z$ and present its contour plot in Fig. \ref{Fig2}(a)
illustrating the variation of $E$ as a function of widths of $m_f = \pm 1$
components. Similarly in Fig. \ref{Fig2}(b), we show the variation of $E$ as 
a function of widths of the $m_f = 0$ component fixing the variational 
parameters $A_1, B, A_2,\sigma_r,$ and $\sigma_z$ at the minimum of energy.
For large widths ($\sigma_r, \sigma_z, \chi_r, \chi_z \to \infty $) in 
Fig. \ref{Fig2} the energy vanishes ($E\to 0$).
The geometric mean of the widths corresponding to the local minima in 
Figs. \ref{Fig2}(a) and (b) is $1.3$ and is close to the  approximate  
width $\sigma$ corresponding to the minimum in Fig. \ref{Fig1} as should be 
the case.

\section{Numerical Procedure}
\label{Sec-III}
The coupled equations  (\ref{gps-1})-(\ref{gps-2}) can be solved by 
time-splitting Fourier pseudo-spectral method \cite{psanand,Wang} and 
time-splitting Crank-Nicolson method \cite{Bao, Muruganandam}. Here, we extend
the Fourier pseudo-spectral method to the coupled GP equations with 
SO-coupling terms and use the same to solve Eqs. (\ref{gps-1})-(\ref{gps-2}). 
The coupled set of GP equations (\ref{gps-1})-(\ref{gps-2}) can be represented
in a simplified form as 
\begin{equation}
\frac{i\partial \Psi}{\partial t} = \left({H_1+H_2+H_3}\right)\Psi\label{GPE},
\end{equation}
where $\Psi=(\psi_{+1},\psi_{0},\psi_{-1})^T$ with $T$ denoting the transpose,
$H_1$, $H_2$ and $H_3$ are $3\times3$ matrix operators defined as
\begin{align}
H_1 &= 
\begin{pmatrix} 
 {\cal H}^-+c_1(n_0+ n_{-}) & 0 &0 \\
0 & {\cal H}+c_1n_{+}&0 \\
0 & 0 & {\cal H}^++c_1(n_0-n_{-}) 
\end{pmatrix},
\end{align}
\begin{align}
H_2 &= 
\begin{pmatrix} 
 0 & c_1\psi_0\psi_{-1}^*&0 \\
c_1\psi_0^*\psi_{-1} & 0 &\psi_0^*\psi_{+1} \\
0 & c_1\psi_0\psi_{+1}^* & 0
\end{pmatrix}, \\
H_3 &=-i\frac{\gamma}{\sqrt{2}} \begin{pmatrix} 
 0 & \partial_-&0 \\
\partial_+ & 0 & \partial_- \\
0 & \partial_+   & 0
\end{pmatrix},
\end{align}
where
\begin{align}
{\cal H}^{\mp}= {\cal H} \mp i\gamma\frac{\partial}{\partial_z},
\quad
n_{\pm}=n_{+1}\pm n_{-1}, \quad  
\end{align}

Now, in the  time-splitting  method  the following equations are solved 
successively
\begin{eqnarray}
\frac{i\partial\Psi}{\partial t} &=& H_1\Psi,\label{GPE1}\\
\frac{i\partial\Psi}{\partial t} &=& H_2\Psi,\label{GPE2}\\
\frac{i\partial\Psi}{\partial t} &=& H_3\Psi\label{GPE3}.
\end{eqnarray}
Equation (\ref{GPE1}) can be numerically solved using Fourier pseudo-spectral 
method 	\cite{Wang} which we employ in this paper or semi-implicit 
Crank-Nicolson method \cite{Muruganandam} and involves additional 
time-splitting of $H_1$ into its spatial derivative and non-derivative parts. 
The numerical solutions of Eq. (\ref{GPE2}) have been discussed in 
Refs. \cite{Wang,Martikainen}.  
We use Fourier pseudo-spectral method to accurately solve Eq. (\ref{GPE3}).
In Fourier space, Eq. (\ref{GPE3}) is
\begin{equation}
\frac{i\partial\widetilde{\Psi}}{\partial t} = 
                              \widetilde{H}_3\widetilde{\Psi}\label{GPEF3},
\end{equation}
where tilde indicates that the quantity has been Fourier transformed.
Hamiltonian $H_3$ in Fourier space is
\begin{equation}
\widetilde{H}_3 = -i\frac{\gamma}{\sqrt{2}} \begin{pmatrix} 
 0 & ik_x+k_y& 0\\
 ik_x-k_y  &0 & ik_x+k_y \\
0 &  ik_x-k_y & 0  
\end{pmatrix}
\end{equation}
The solution of Eq. (\ref{GPEF3}) is
\begin{align}
\widetilde{\Psi}(t+dt) &= e^{-i\widetilde{H}_3 dt} 
\widetilde{\Psi}(t) = e^{-i\hat{O}} \widetilde{\Psi}(t),\\
	   &= \left(I + \frac{\cos{\Omega}-1}{\Omega^2}\hat{O}^2-
i\frac{\sin{\Omega}}{\Omega}\hat{O}\right)\widetilde{\Psi}(t),\label{GPEF3S}
\end{align}
where $\Omega = \sqrt{|A|^2 +|B|^2}$, where 
$A = -i\frac{\gamma}{\sqrt{2}}\left(ik_x + k_y\right)dt$ 
and $ B = -i\frac{\gamma}{\sqrt{2}}\left(ik_x -k_y\right)dt$, 
and $\hat{O}$ is defined as
\begin{equation}
\hat{O} = \begin{pmatrix}
0 & A & 0\\
A^*&0& B^*\\
0&B&0
\end{pmatrix}.
\end{equation}
The wavefunction in Eq. (\ref{GPEF3S}) is in Fourier space and can be inverse 
Fourier transformed to obtain the solution  in configuration  space.
In this study, in  space and time discretizations, we use space and time 
steps of $0.1$ and $0.005$, respectively, in imaginary-time simulation, 
whereas in real-time simulation these are, respectively, $0.1$ and $0.0005$.
Numerically, the component wave functions of a stationary 
$(-1,0,+1)$ vortex-bright  soliton are calculated by an imaginary-time 
propagation of Eqs. (\ref{gps-1}) and (\ref{gps-2}) using the initial guess 
of component wave functions (\ref{ansatz1}) and (\ref{ansatz2}).

\begin{figure}[t]
\begin{center}
\includegraphics[trim = 0mm 0mm 0mm 0mm, clip,width=0.49\linewidth,clip]{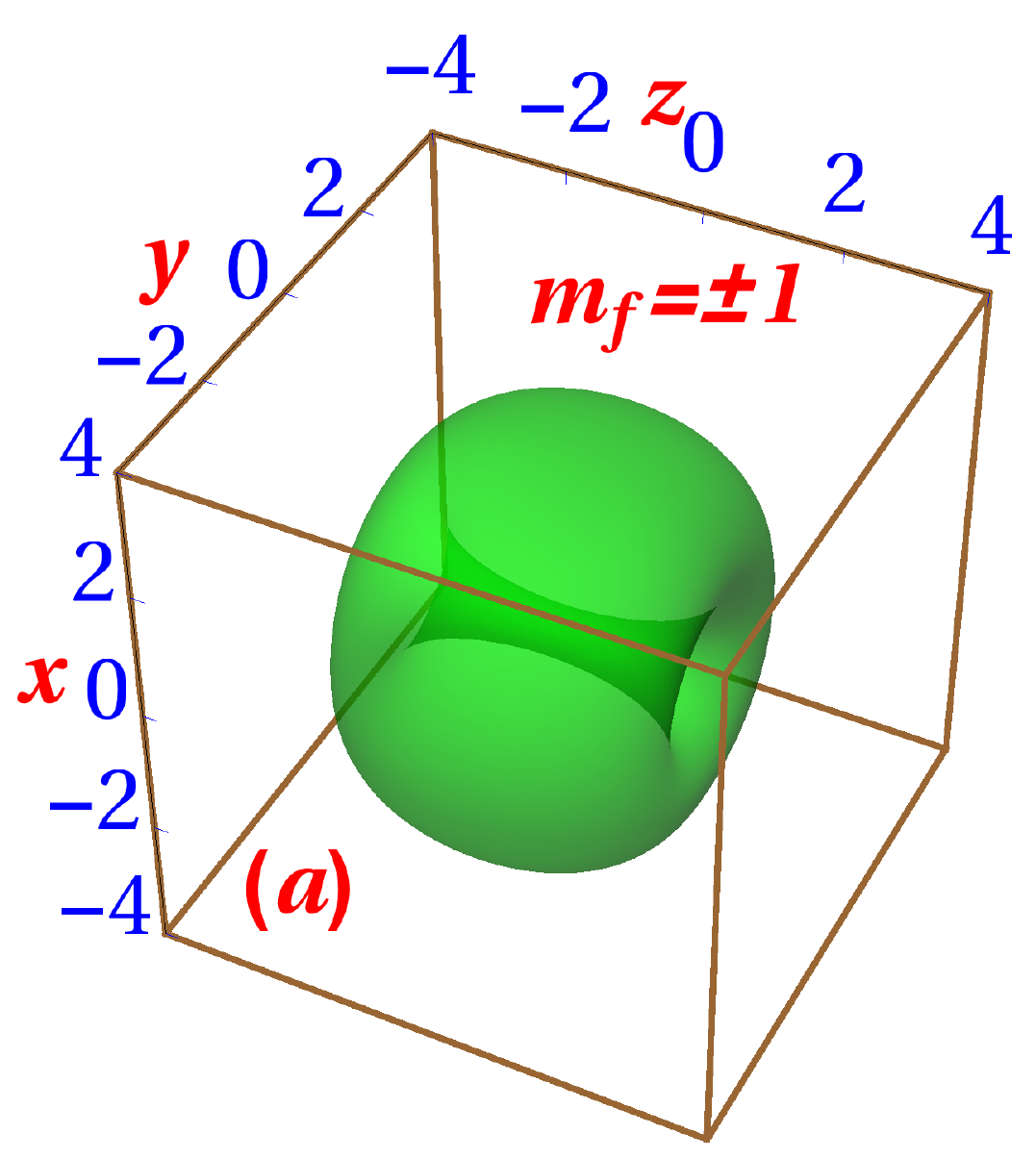}
\includegraphics[trim = 0mm 0mm 0mm 0mm, clip,width=0.49\linewidth,clip]{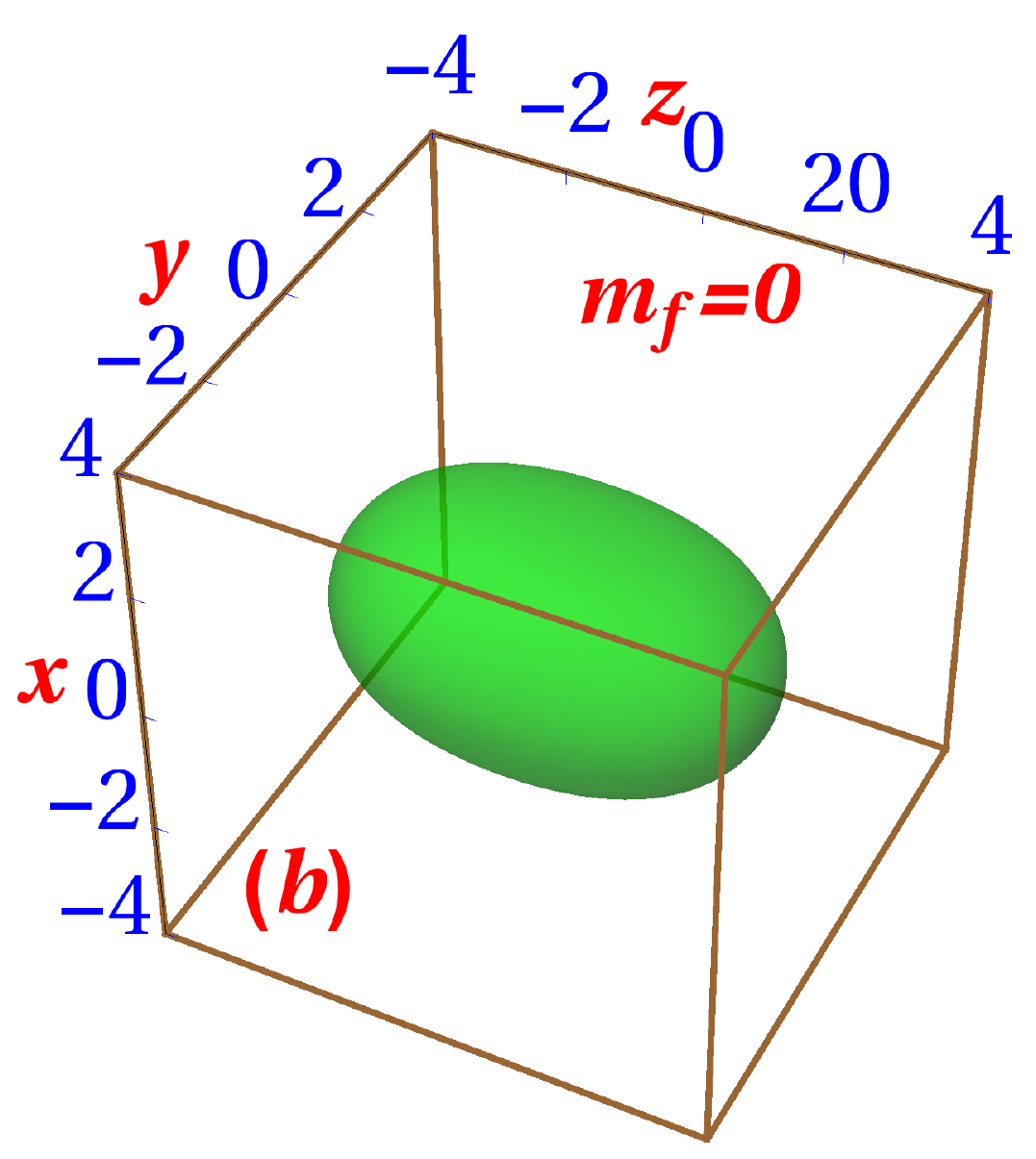}\\
\includegraphics[trim = 0mm 0mm 0mm 0mm, clip,width=0.6\linewidth,clip]{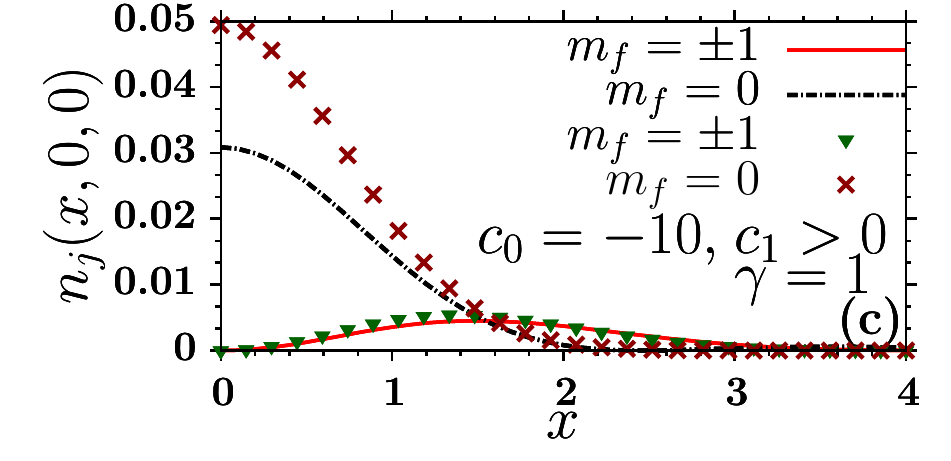}
\includegraphics[trim = 0mm 0mm 0mm 0mm, clip,width=0.38\linewidth,clip]{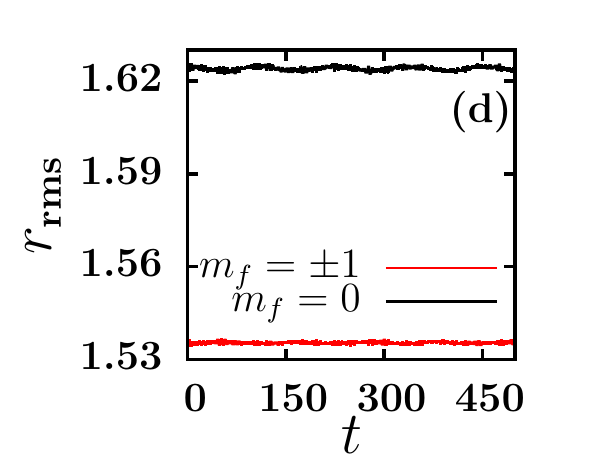}\\
\includegraphics[trim = 0mm 0mm 0mm 0mm, clip,width=.34\linewidth,clip]{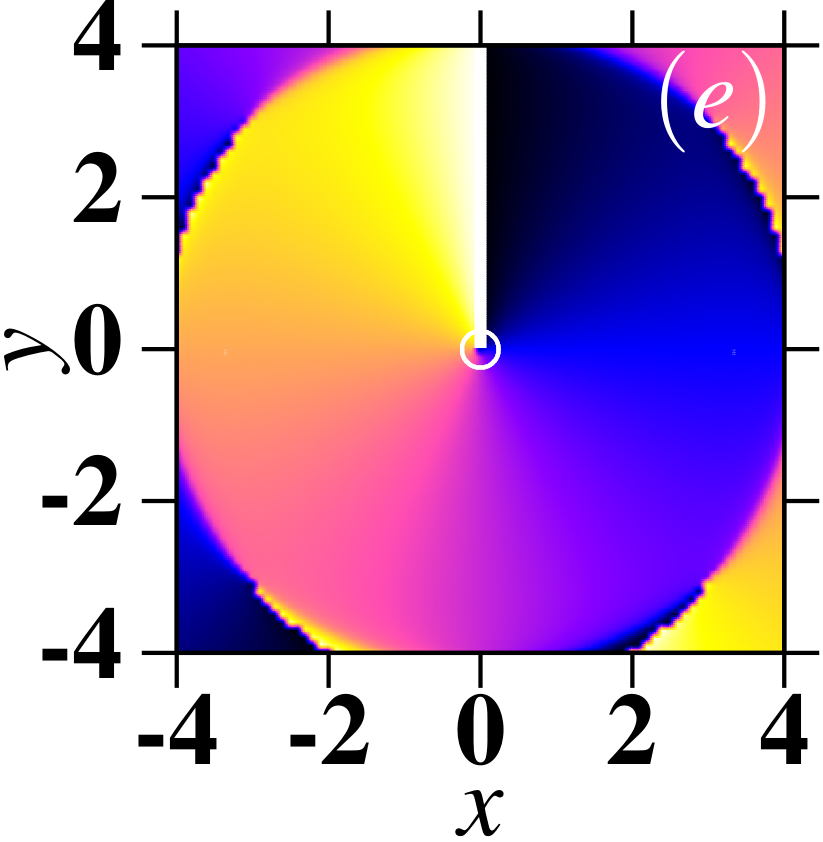}
\includegraphics[trim = 0mm 0mm 0mm 0mm, clip,width=.285\linewidth,clip]{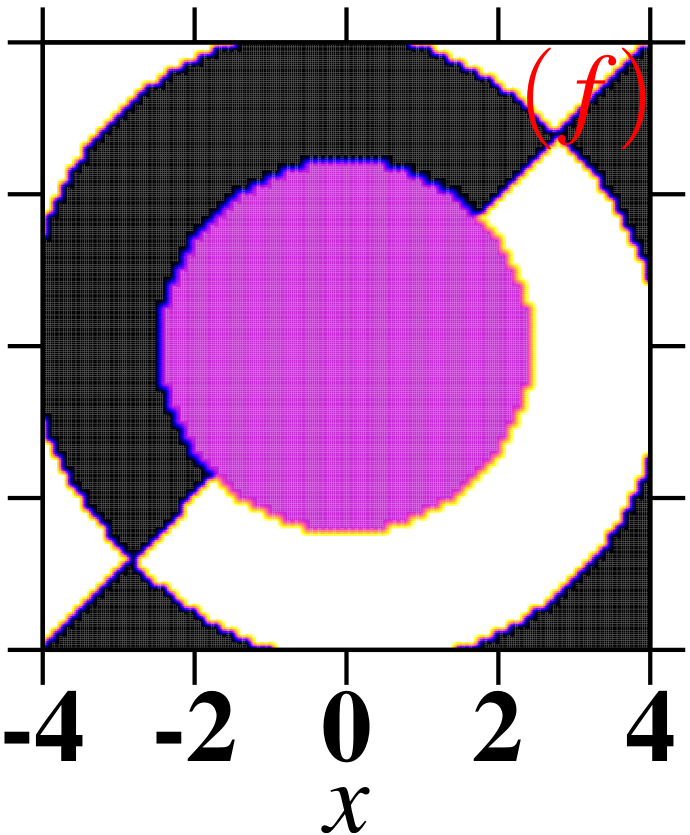}
\includegraphics[trim = 0mm 0mm 0mm 0mm, clip,width=.35\linewidth,clip]{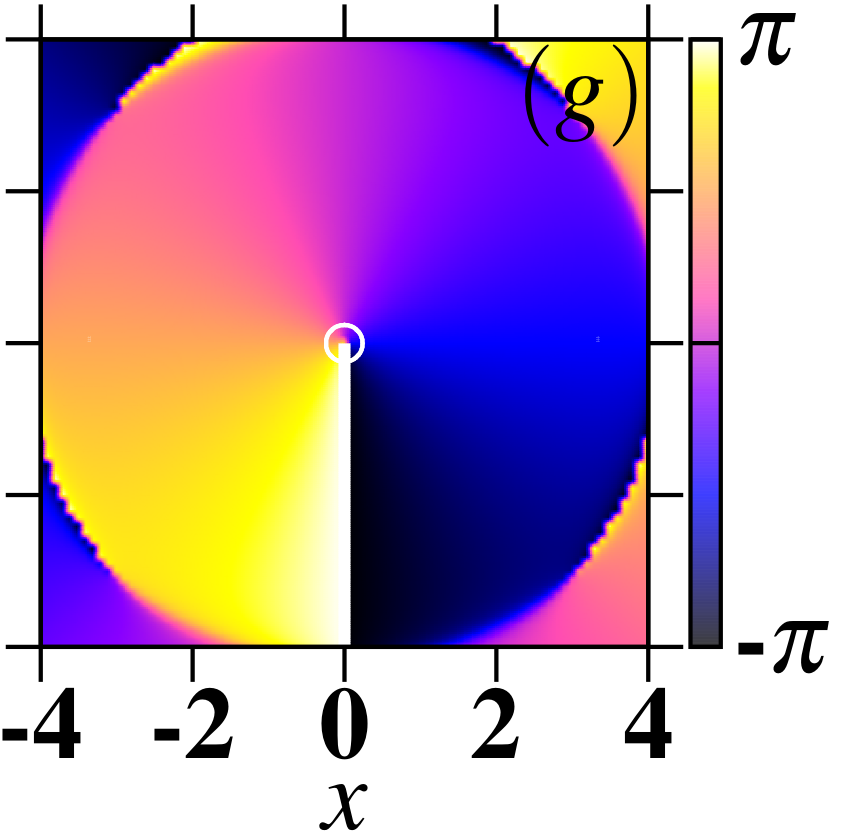}
\caption{(a)  Numerical isodensity contour of $|\psi_{m_f}|^2$
 for $m_f = \pm 1$ components for  $c_0 = -10, c_1 \ge 0$, and  $\gamma = 1$ 
 in an axisymmetric $(- 1,0,+ 1)$ vortex-bright soliton, (b) the same for the 
 $m_f  = 0$ component. The density on the contour is 0.0007.
(c) Numerical 
 (line) and variational (chain of symbols) results for 
radial densities in the $z=0$ plane $n_j(x,0,0)$ versus $x$.   (d) Numerical result of rms 
 sizes of the component wave functions versus time as obtained in real-time 
 simulation of the solution shown in (a)-(c); upper-black(dark)  and lower-red(light) curves correspond to rms sizes of $m_f= 0$ and
$m_f = \pm 1$ components, respectively. All quantities in this and 
 following figures are dimensionless. (e), (f) and (g) show the phase distribution of $m_f=+1$, $m_f=0$, and $m_f=-1$ components, 
respectively, in $z=0$ plane.
The norms $\int n_{j}d{\bf r}$ of three components obtained
numerically are 0.29, 0.42, and 0.29, for $j=1,0,$ and $-1$, respectively.}
\label{Fig3}
\end{center}
\end{figure}

\section{Numerical results}
\label{Sec-IV}
\subsection{Axisymmetric vortex-bright soliton}
\label{Sec-IVA} 
First we consider the  $(- 1,0,+ 1)$ vortex-bright soliton in the polar domain.
The numerical results for surfaces of constant density (isodensity contour) 
in coordinate space for an axisymmetric $(- 1,0,+ 1)$ vortex-bright soliton  
with $c_0 = -10, c_1 \ge 0$, and $ \gamma = 1$ are shown in Figs. 
\ref{Fig3}(a)-(b). In the polar domain, $c_1>0$, the results for 
$(- 1,0,+ 1)$ vortex-bright solitons are independent of the parameter $c_1$.
To compare the numerical and variational results, we show in 
Fig. \ref{Fig3}(c)  the numerical and variational densities in the radially outward direction $n_j(x,0,0)$ 
versus $x$ in the $z = 0$ plane. In the $z = 0$ plane, the densities 
of the three components have the oscillating asymptotic behavior given by $ n_{\pm 1}(x,y,0)\sim J_1(k_\rho \rho)^2$ 
and $n_{0}(x,y,0) \sim J_0(k_\rho \rho)^2$ as discussed in Sec. \ref{Sec-IIA}. The Gaussian
ansatz does not capture the oscillating tail properly resulting in the difference
in numerical and variational results in Fig. \ref{Fig3}(c).  
To demonstrate the dynamical stability of the 
vortex-bright soliton, we performed real-time simulation of the 
imaginary-time profile as the initial state over a long interval of time.  
The dynamical stability in real time propagation confirm a stable ground state or a metastable state.
The steady oscillation of the root-mean-square (rms) size 
($r_{\mathrm{rms}}\equiv \sqrt{x_{\mathrm{rms}}^2 + 
y_{\mathrm{rms}}^2+z_{\mathrm{rms}}^2}$ ~\ ) of the components as shown in 
Fig. \ref{Fig3} (d), corresponding to the $(- 1,0,+ 1)$ vortex-bright soliton
shown in Figs. \ref{Fig3} (a)-(c), demonstrates the  dynamical stability of the soliton.
The nature of phase singularity, if any, in the components
can be inferred from the phase plots of three components in $z = 0$ plane 
as are shown in Figs. \ref{Fig3}(e)-(g), where the location of the vortex is indicated by a 
small circle in the phase plots.

\begin{figure}[t]
\begin{center}
\includegraphics[trim=0mm 0mm 0mm 0mm,clip,width=0.45\linewidth,clip]{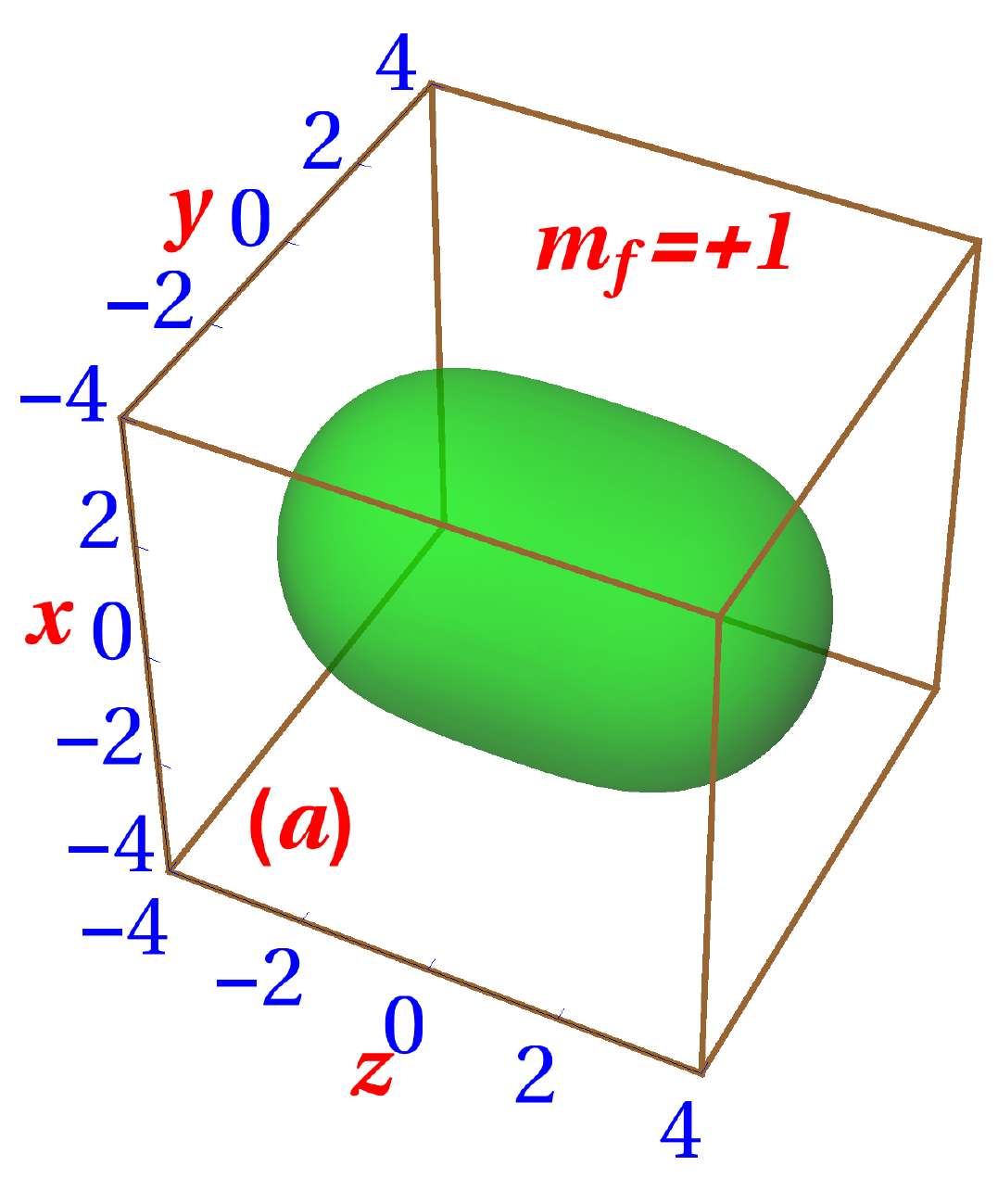}
\includegraphics[trim=0mm 0mm 0mm 0mm,clip,width=0.45\linewidth,clip]{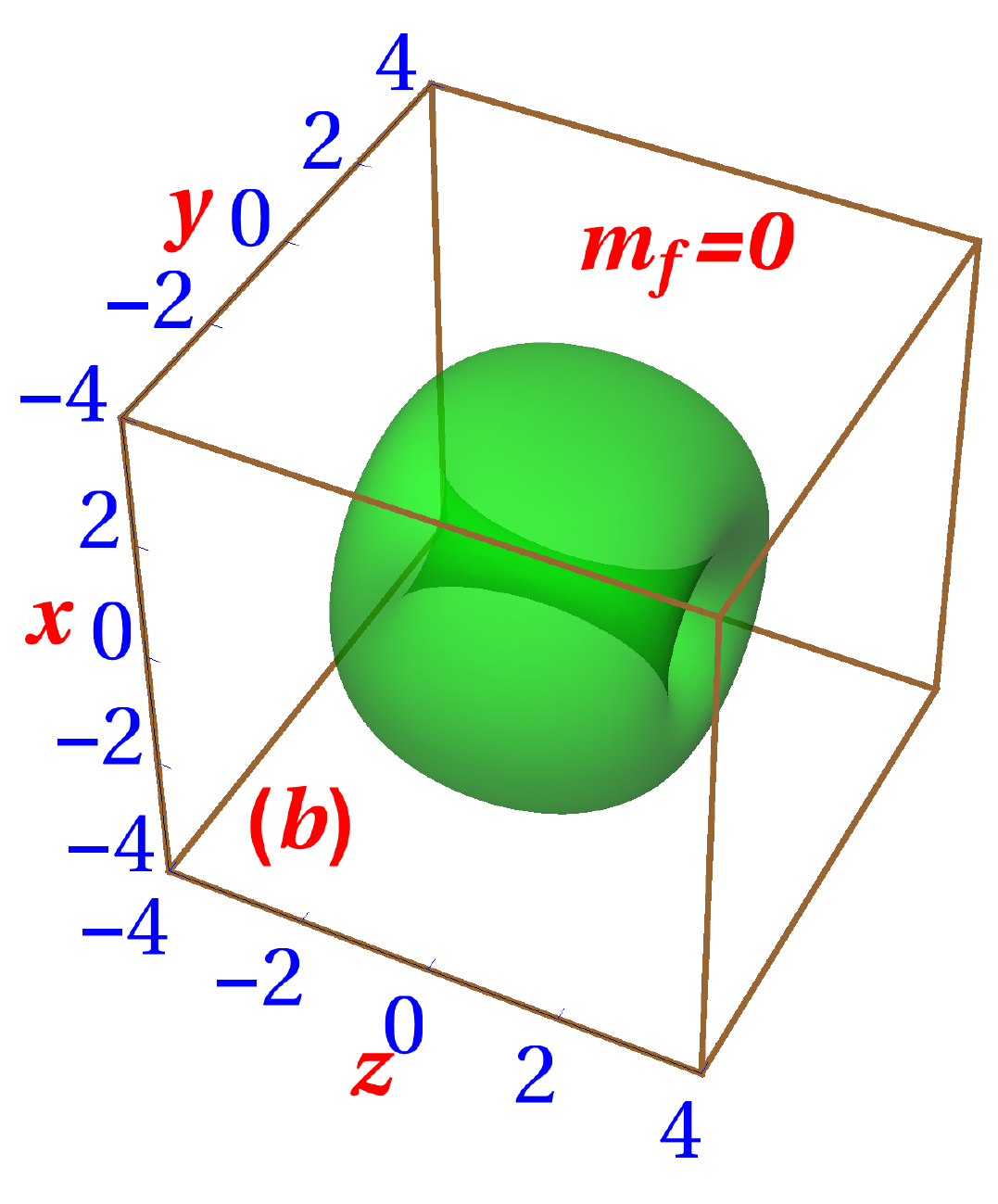}
\includegraphics[trim=0mm 0mm 0mm 0mm,clip,width=0.45\linewidth,clip]{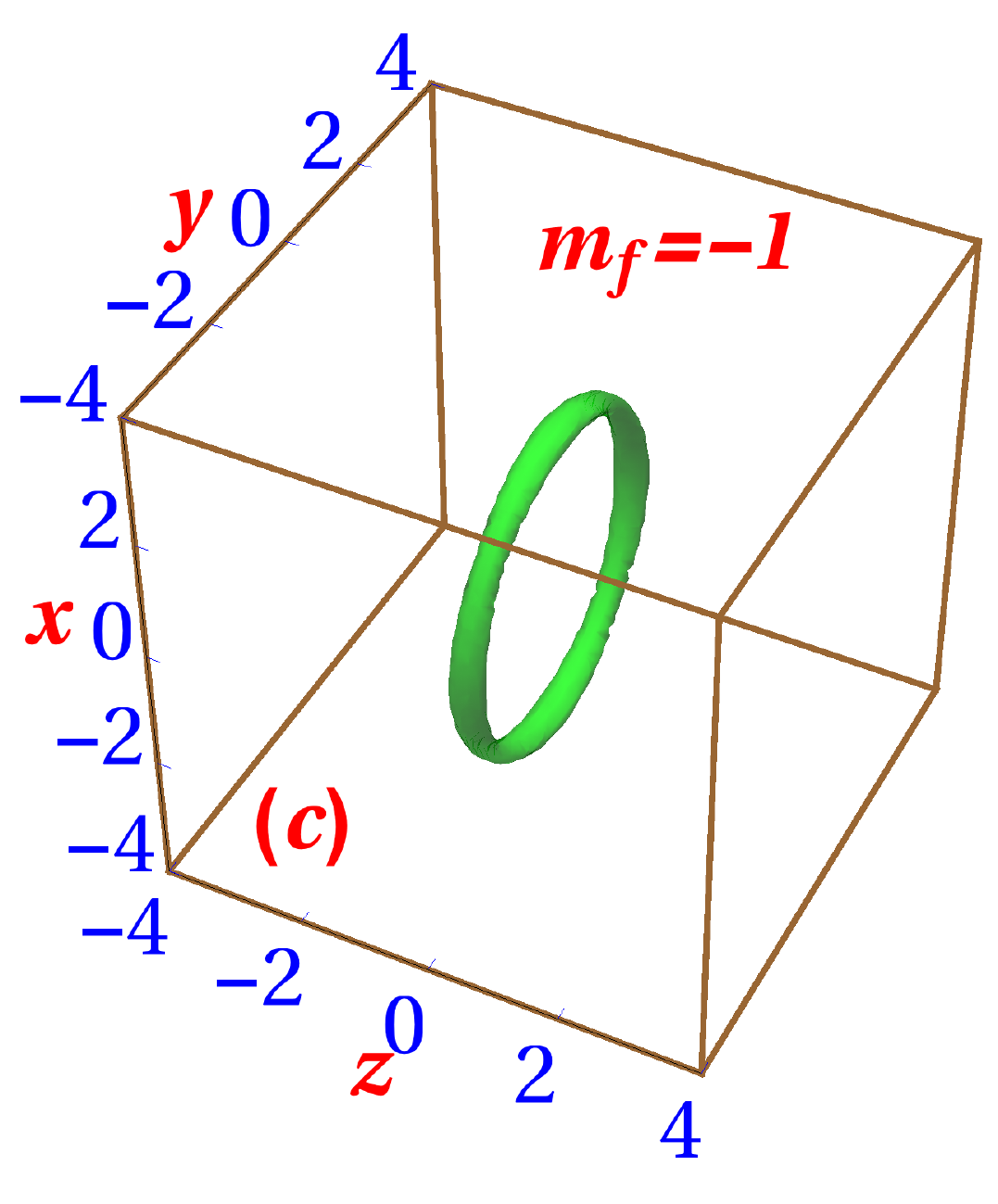} 
\includegraphics[trim=0mm 0mm 0mm 0mm,clip,width=0.48\linewidth,clip]{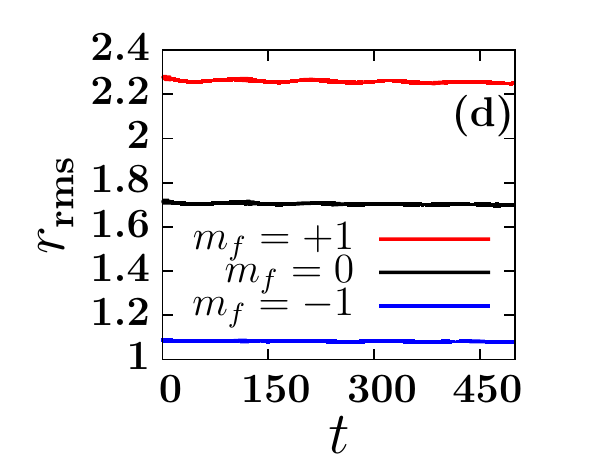}\\ 
\includegraphics[trim = 0mm 0mm 0mm 0mm, clip,width=.34\linewidth,clip]{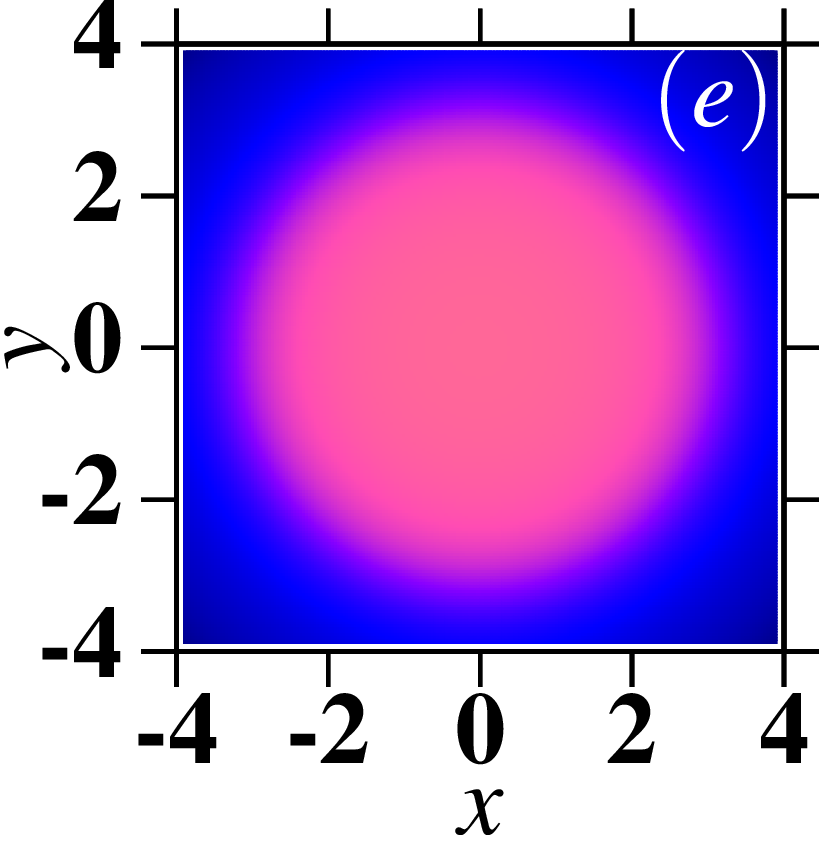}
\includegraphics[trim = 0mm 0mm 0mm 0mm, clip,width=.285\linewidth,clip]{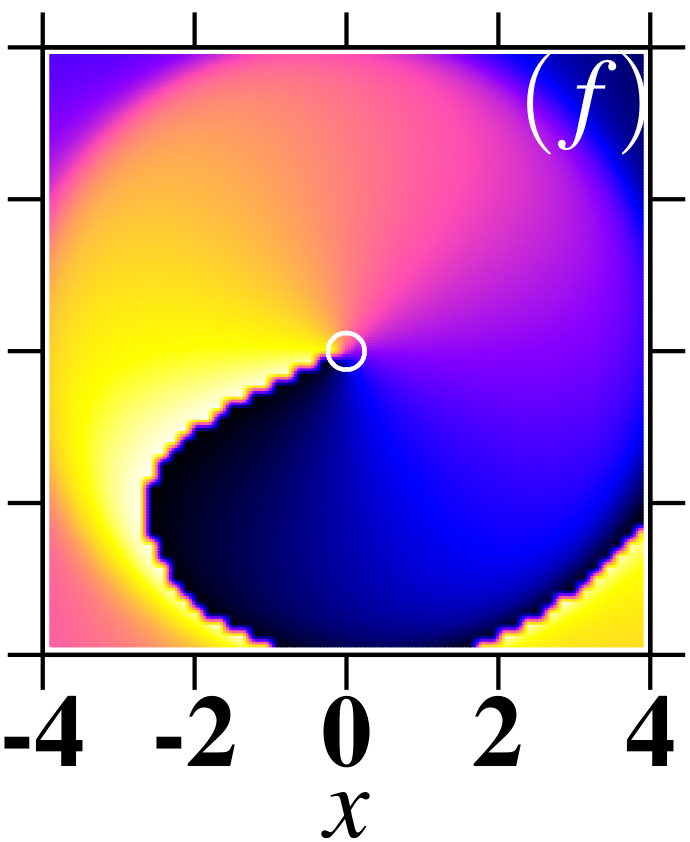}
\includegraphics[trim = 0mm 0mm 0mm 0mm, clip,width=.345\linewidth,clip]{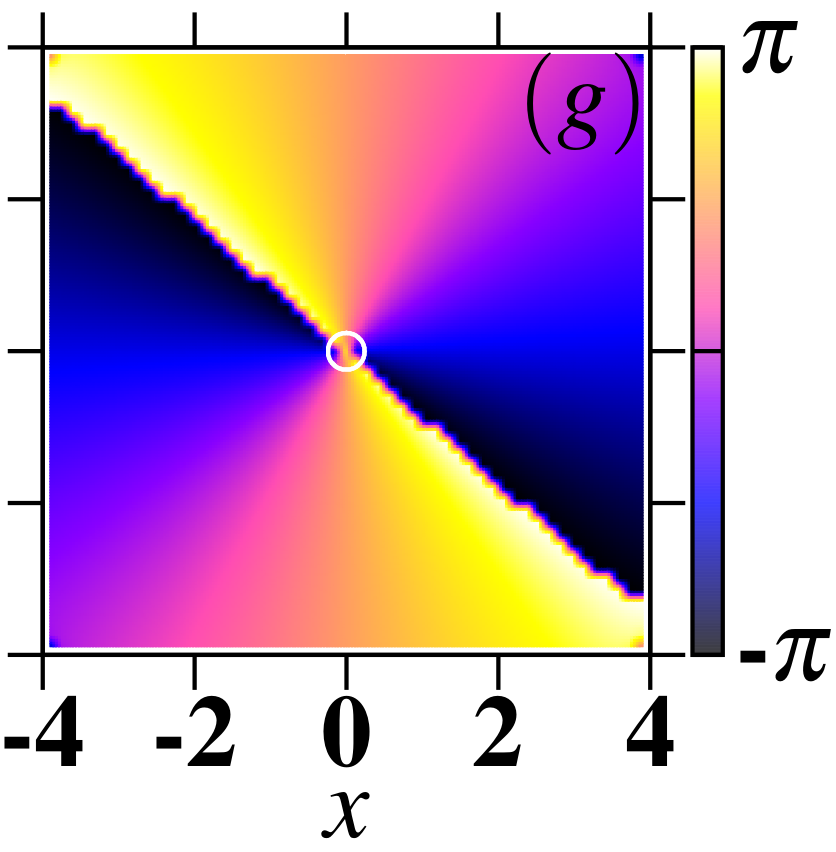}
\caption{ Numerical isodensity contour of $|\psi_{m_f}|^2$ for (a) $m_f = +1$, 
(b) $m_f = 0$, and (c) $m_f = -1$ components of a polar $(0,+1,+2)$ 
vortex-bright soliton for $c_0=-10, c_1= 0.1$ and $\gamma=1$. The density on 
the contour is 0.0007. 
The isosurfaces are translucent to show the presence of
vortex lines. (d) Numerical result of rms size ($r_{\mathrm{rms}}$) of the 
component wave functions 
versus time as obtained in real-time simulation of 
the solution shown in (a)-(c); upper-red, middle-black and lower-blue curves correspond 
to $m_f = +1$, $m_f = 0$, and $m_f = -1$ components, respectively. 
(e), (f) and (g) show the phase distribution of $m_f=+1$, $m_f=0$, and $m_f=-1$ components, respectively, in $z=0$ plane.
The norms of the three components are
0.63, 0.29, 0.09 for $m_f=+1$, $m_f=0$, and $m_f=-1$ respectively.
}
\label{Fig4}
\end{center}
\end{figure}

The numerical isodensity contour of the axisymmetric $(0,+ 1,+ 2)$ 
vortex-bright soliton for  $c_0=-10$, $\gamma=1$ and $c_1 = 0.1$ are shown in 
Figs. \ref{Fig4}(a)-(c). The results for the axisymmetric $(0,+ 1,+ 2)$ 
vortex-bright soliton  in the polar domain depend on the value of $c_1$.
The numerical result  is obtained by an imaginary-time simulation of 
Eqs. (\ref{gps-1}) and (\ref{gps-2}) with the initial guess of component 
wave functions  $\psi_1^{\mathrm{inital}} \approx \psi_{\rm Gauss},~ 
\psi_0^{\mathrm{inital}} \approx (x+iy)\psi_{\rm Gauss},~
\psi_{-1}^{\mathrm{inital}}\approx(x+iy)^2\psi_{\rm Gauss}$, where 
$\psi_{\rm Gauss}$ is a Gaussian wavefunction. The $(0,+1,+2)$ vortex-bright 
soliton shown in \ref{Fig4}(a)-(c) has higher energy than the $(-1,0,+1)$ 
soliton shown in Figs. \ref{Fig3}(a)-(c). Again, the $(0,+1,+2)$ vortex-bright
soliton is dynamically stable as is demonstrated by the steady oscillation of 
rms sizes of the component wavefunctions in Fig. \ref{Fig4}(d). 
We do not observe any decay of the angular momentum in each component in
both $(-1,0,+1)$ and $(0,+1,+2)$ vortex-bright solitons. The charge of the
vortices in the three components is evident from the phase-profile of three
components shown in Figs. \ref{Fig4}(g)-(f) with a small circle used
to indicate the exact location of phase singularity.

\begin{figure}[t]
\begin{center}
\includegraphics[trim=0cm 0cm 0cm 0cm,clip,width=0.45\linewidth,clip]{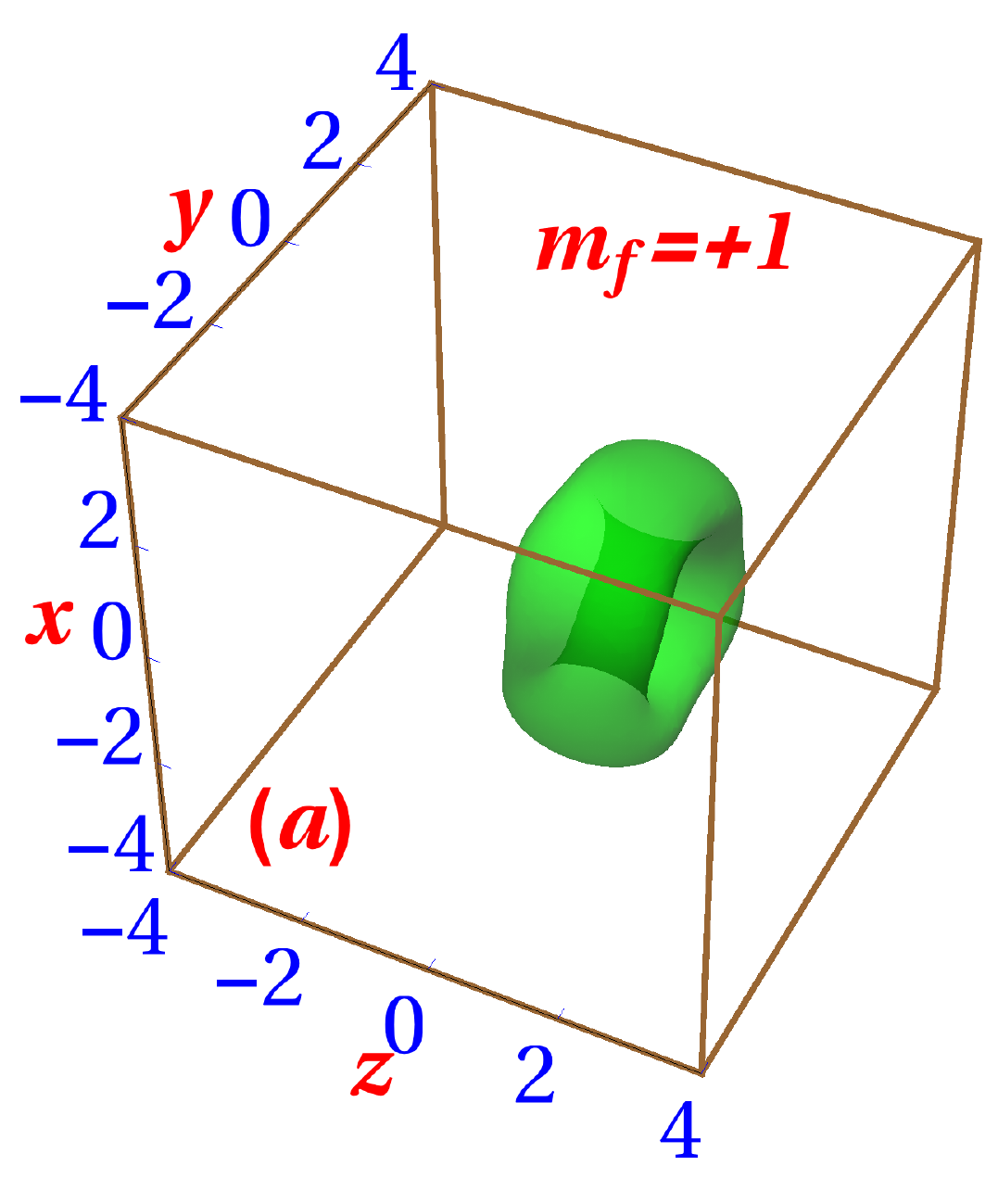}
\includegraphics[trim=0cm 0cm 0cm 0cm,clip,width=0.45\linewidth,clip]{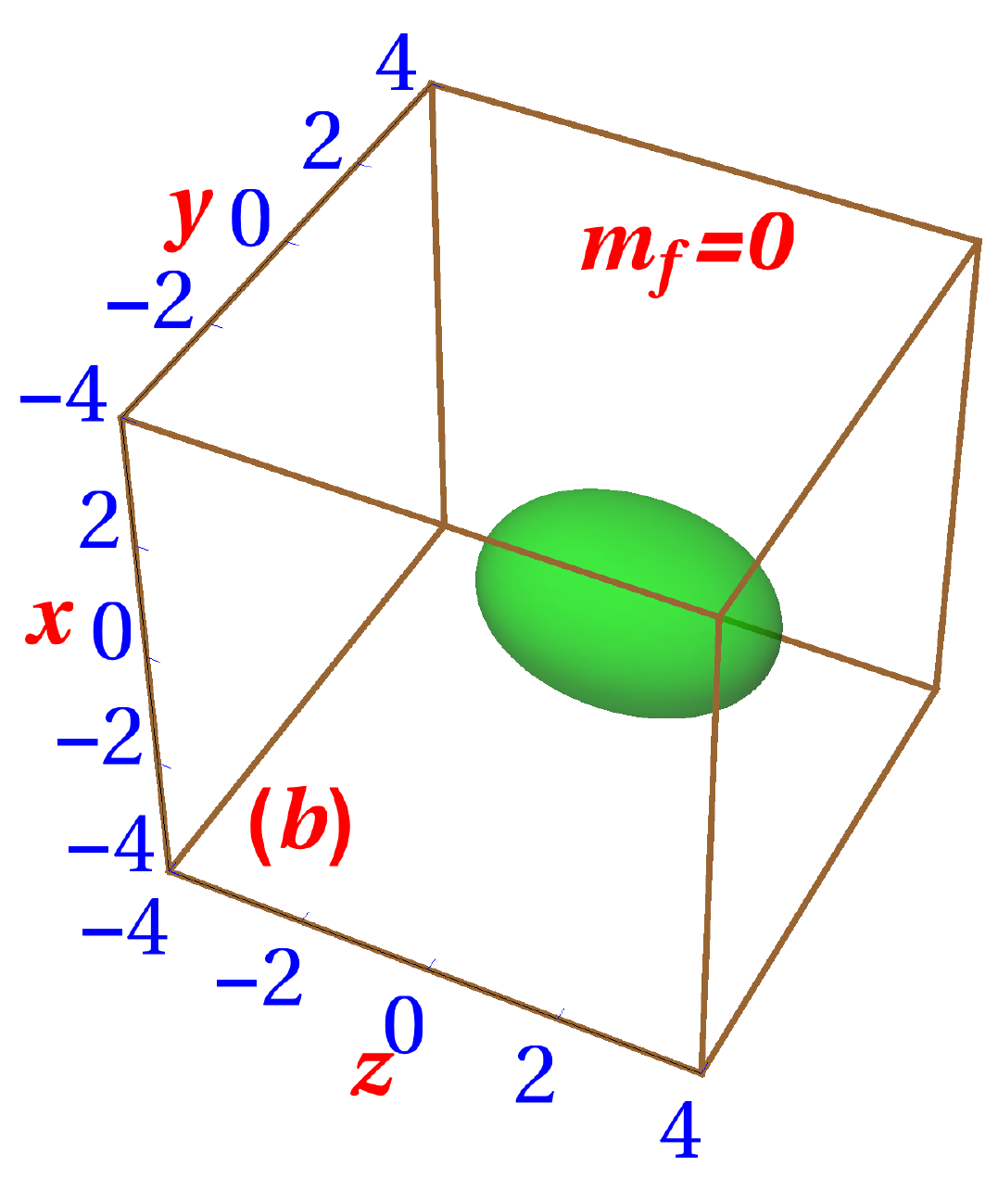} 
\includegraphics[trim=0cm 0cm 0cm 0cm,clip,width=1\linewidth,clip]
               {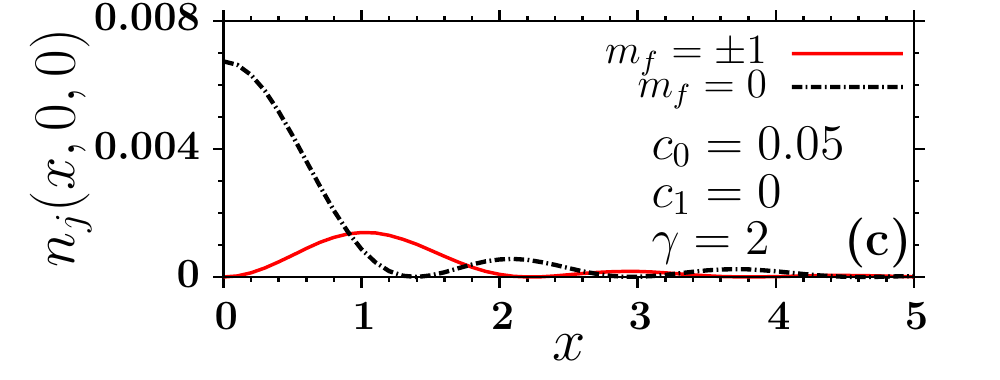} 
\includegraphics[trim=0.cm 0cm 0.cm 0.cm,clip,width=1\linewidth,clip]
               {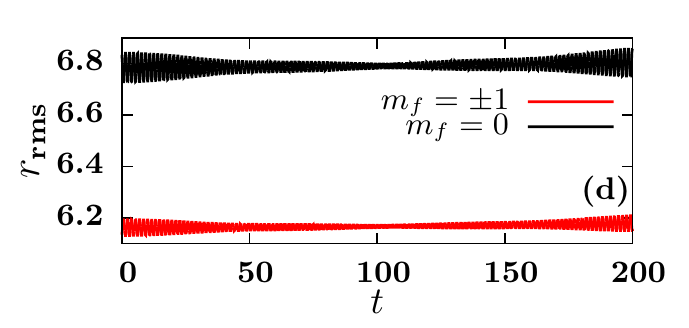} 
\caption{Numerical isodensity contour of $|\psi_{m_f}|^2$ for (a) 
$m_f = \pm 1$, (b) $m_f = 0$, components of an axisymmetric (-1,0,+1) soliton 
with $c_0 = 0.05$, $c_1 = 0$ and $\gamma =2$. The density on the contour is 
0.001. (c)  Numerical   results for 
radial densities in the $z=0$ plane $n_j(x,0,0)$ versus $x$.   (d) Numerical result of rms sizes of the 
component wave wavefunctions versus time as obtained in real-time simulation 
of the solution shown in (a)-(c); upper-black(dark) and lower-red(light) curves correspond to $m_f = 0$ and $m_f = \pm 1$ components, respectively. 
The norms of the three
components are 0.30, 0.39, 0.30, for $m_f=+1,0$, and $m_f=-1$ components, 
respectively.}
\label{Fig5}
\end{center}
\end{figure}

Solitons are also possible in a repulsive spinor BEC. Using variational 
approximation in Sec. \ref{Sec-IIB}, we found that the SO-coupled BEC may
have stable vortex bright soliton for $c_0\ge 0$, viz. Fig. \ref{Fig1}. 
The numerical isodensity contour of such a (-1,0,+1) vortex-bright soliton 
with $c_0 = 0.05, c_1 \ge 0$, and $\gamma = 2$ are shown in 
Figs. \ref{Fig5}(a)-(b). Although a larger $\gamma$ leads to easier bright-soliton formation, viz. Fig. \ref{Fig1}, 
we could not use a much larger $\gamma$ because of the numerical difficulty in treating a large undulating tail in density \cite{str},
associated with a large $\gamma$.  
In asymptotic region in the $xy$ plane, the densities
of the $m_f=\pm1$ and $m_f= 0$ components vary as $\sim \sin(k_\rho \rho-\pi/4)$
and $\sim \cos(k_\rho \rho-\pi/4)$, respectively, viz. Sec. \ref{Sec-IIA} leading to the oscillating tail with period
inversely dependent on $k_\rho = \sqrt{\gamma^2 - k_z^2}$.
The long-time
oscillation in the  rms radius $r_{\rm rms}$ of the three components, corresponding
to the $(- 1,0,+ 1)$  vortex-bright soliton shown in Figs. \ref{Fig5}(a)-(c), as obtained upon real-time evolution of the imaginary-time profile exhibited  
 in Fig.  \ref{Fig5}(d), establishes the  dynamical stability of the soliton.  


A finite space step  $\Delta x$ and a finite space domain $L$ of discretization 
along the corresponding direction set a 
limit on the accuracy of the  numerical scheme. The typical wavelength of spatial modulations of 
density  should be much larger than the space step for an adequate 
resolution and for obtaining reliable results $-$ a  
condition which is very well satisfied 
for the SOC values considered in the present work. Larger SOC values will require 
even  smaller space step  to adequately treat the high-wavenumber spatial modulations of 
density. For example, comparing the density plots of Figs. 3(c) and 5(c), we find that a larger SOC ($\gamma=2$) 
in the latter has led to  pronounced spatial modulations of density  compared to the former
with a smaller SOC ($\gamma =1$). The use of the same spatial step ($\Delta x = 0.1$) in both cases has set a limit on the 
accuracy in Fig. 5(c), which is responsible for the rapid  temporal oscillation of the rms size in Fig. 5(d). Also, the space domain 
$L$ should be much larger than the size of the  condensate along that direction for obtaining accurate results.  
However, any finite $L$ leads to the slow temporal oscillation of the rms size in Fig. 3(d), which is also present in  Fig. 5(d) as a modulating envelope over 
the rapid oscillation. As the space domain $L$ is increased these oscillations with smaller frequency 
will tend to vanish. The use of  small space step  will reduce the fast temporal 
oscillations of the rms size in Fig. 5(d), whereas larger space domain will suppress the slow 
oscillation of the rms size.


\subsection{Asymmetric solitons}
\label{Sec-IVB}

\begin{figure}[t]
\begin{center}
\includegraphics[trim=0cm 0cm 0cm 0cm,clip,width=0.45\linewidth,clip]{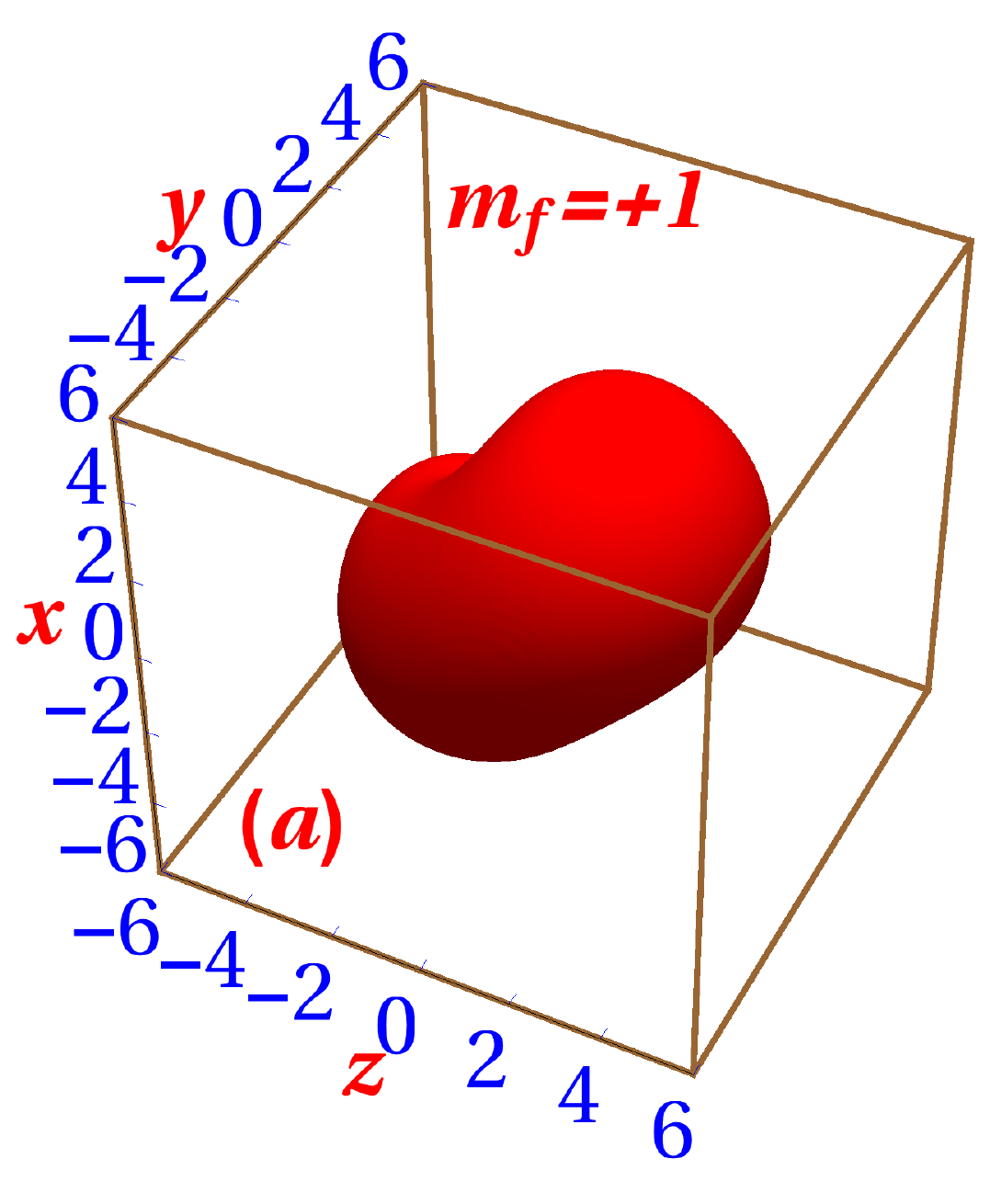}
\includegraphics[trim=0cm 0cm 0cm 0cm,clip,width=0.45\linewidth,clip]{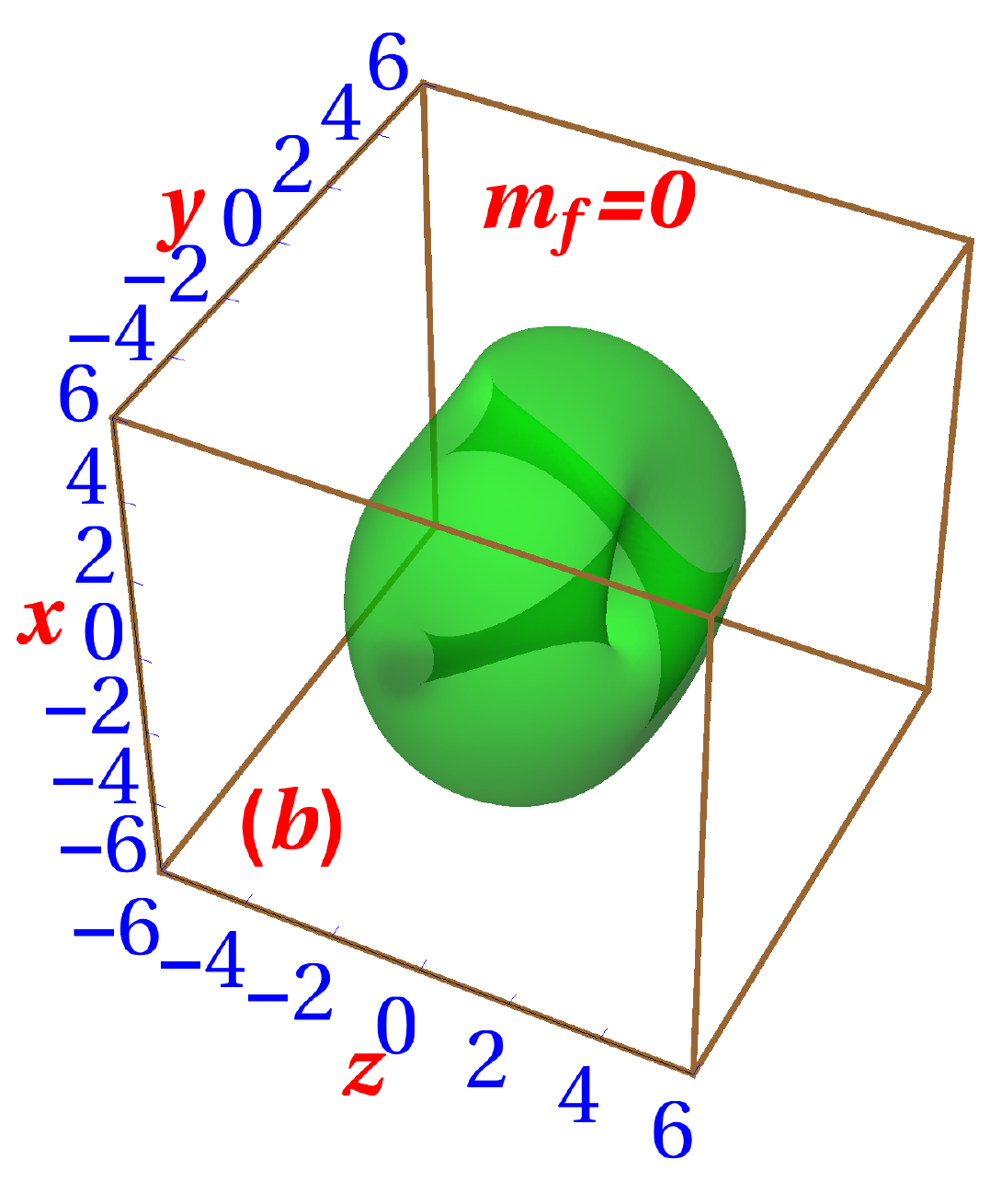}
\includegraphics[trim=0cm 0cm 0cm 0cm,clip,width=0.45\linewidth,clip]{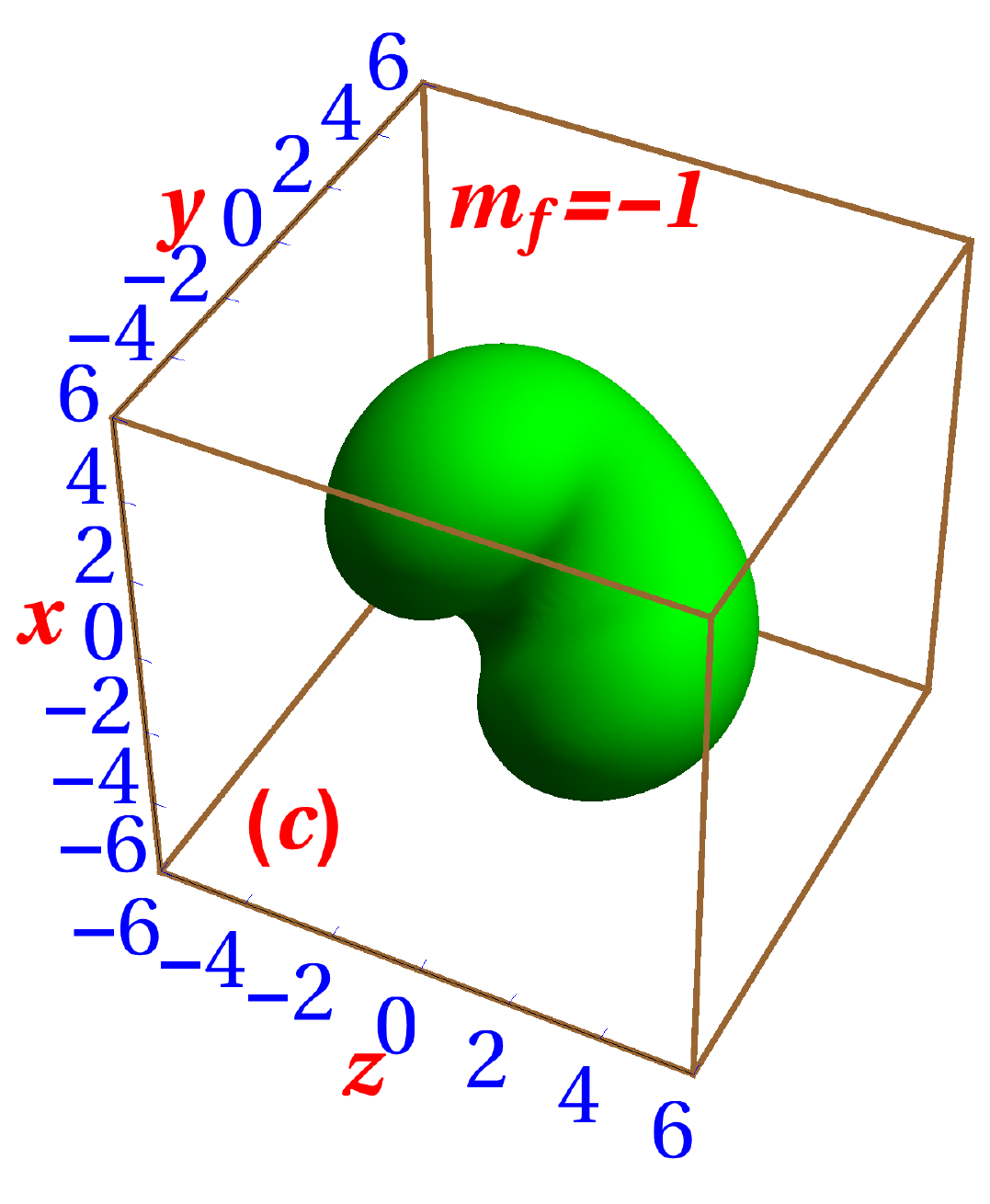}
\includegraphics[trim=0cm 0cm 0cm 0cm,clip,width=0.45\linewidth,clip]{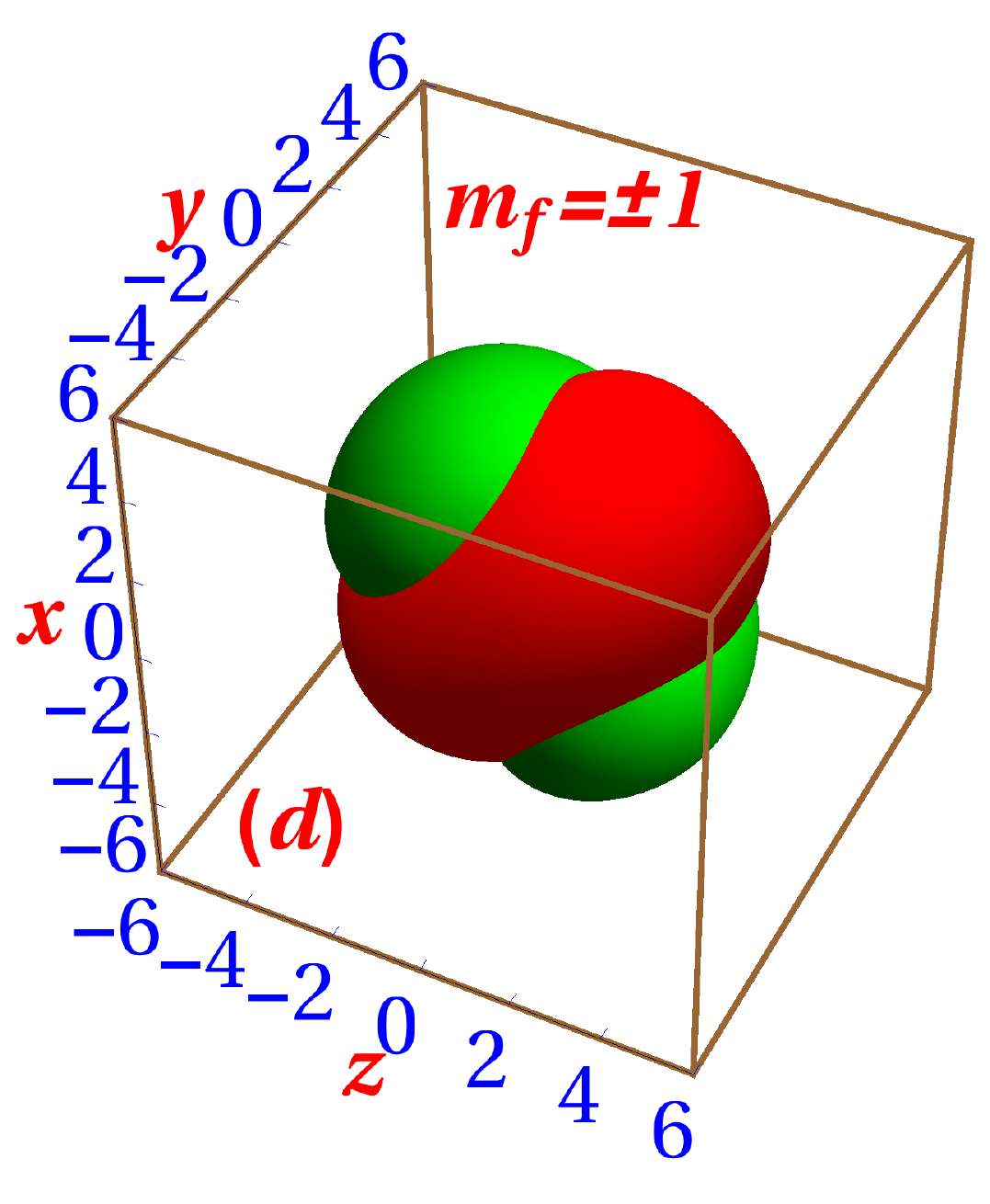}
\caption{Numerical isodensity contour of $|\psi_{m_f}|^2$ for (a) $m_f = +1$, 
(b) $m_f = 0$, (c) $m_f = -1$ components of an  asymmetric  soliton with 
$c_0 = -10$, $c_1 = -1$ and $\gamma =1$. The density on the contour is 0.00008.
(d) shows the isosurfaces in (a) and (c) together. The norms of three components are $0.32$, $0.36$, $0.32$ for $m_f = +1, 0$, and $m_f=-1$, respectively.}
\label{Fig6}
\end{center}
\end{figure}

\begin{figure}[t]
\begin{center}
 \begin{tabular}{lll}
\includegraphics[trim = 0cm 0cm 0cm 0cm, clip,width=0.35\linewidth,clip]{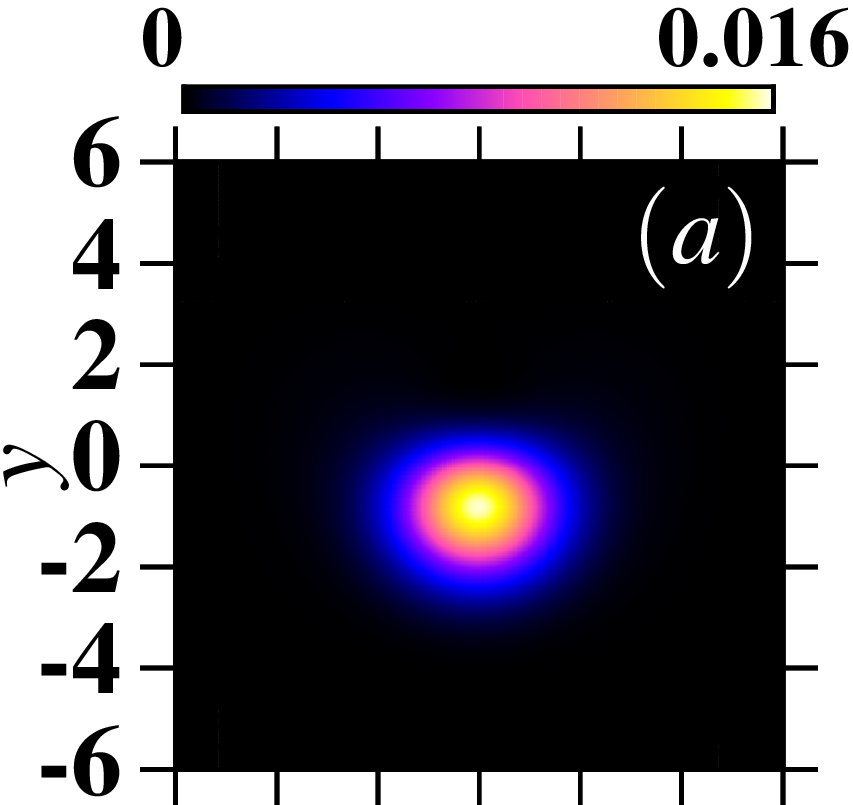}
\includegraphics[trim = 0cm 0cm 0cm 0cm, clip,width=0.285\linewidth,clip]{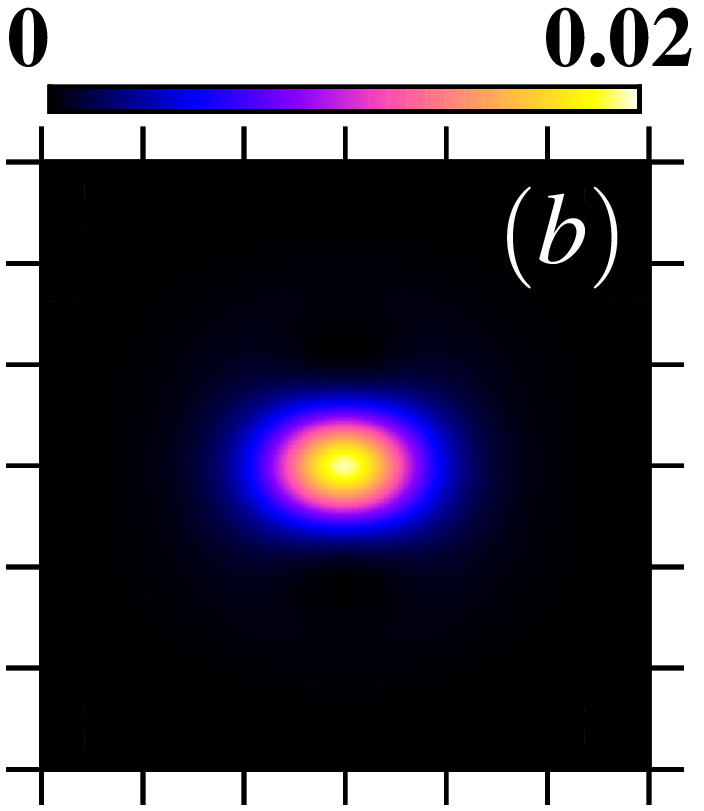}
\includegraphics[trim = 0cm 0cm 0cm 0cm, clip,width=0.295\linewidth,clip]{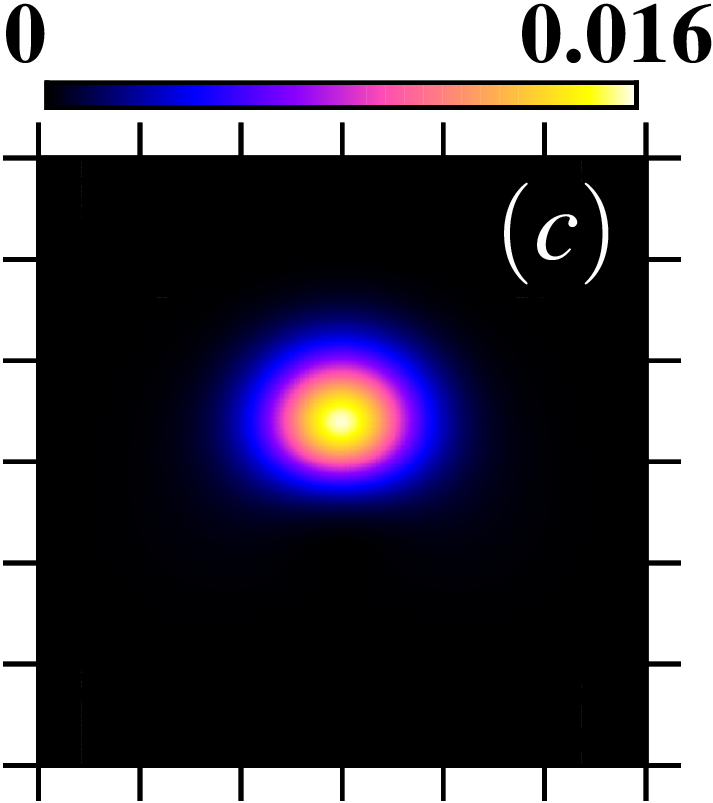}\\
\includegraphics[trim = 0cm 0cm 0cm 0cm, clip,width=0.34\linewidth,clip]{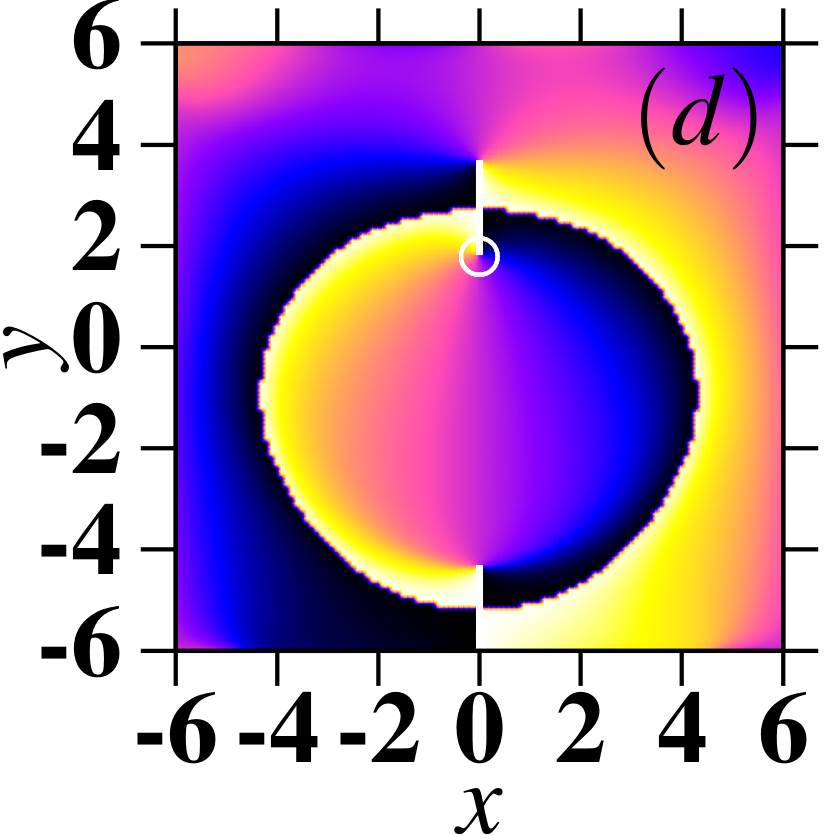}
\includegraphics[trim = 0cm 0cm 0cm 0cm, clip,width=0.285\linewidth,clip]{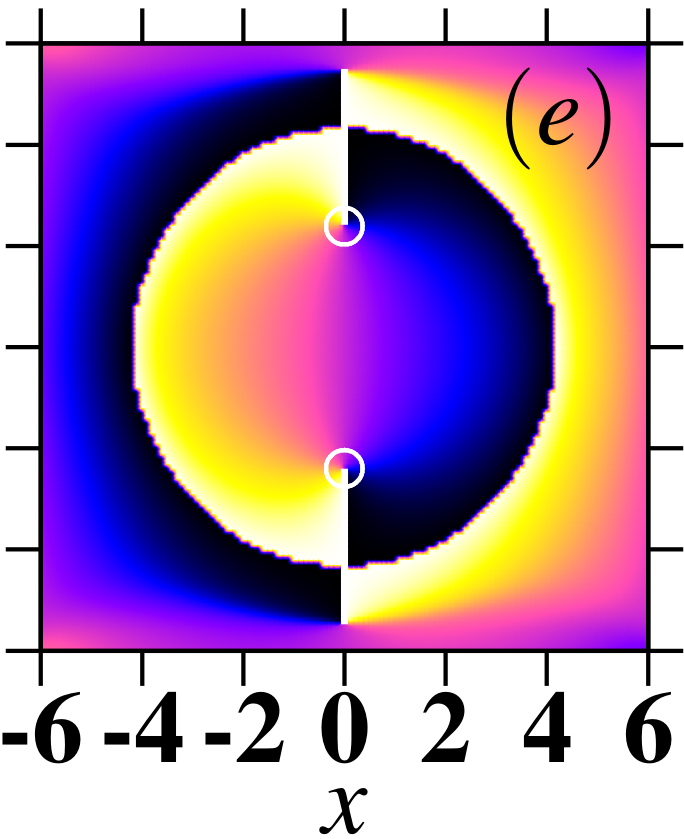}
\includegraphics[trim = 0cm 0cm 0cm 0cm, clip,width=0.345\linewidth,clip]{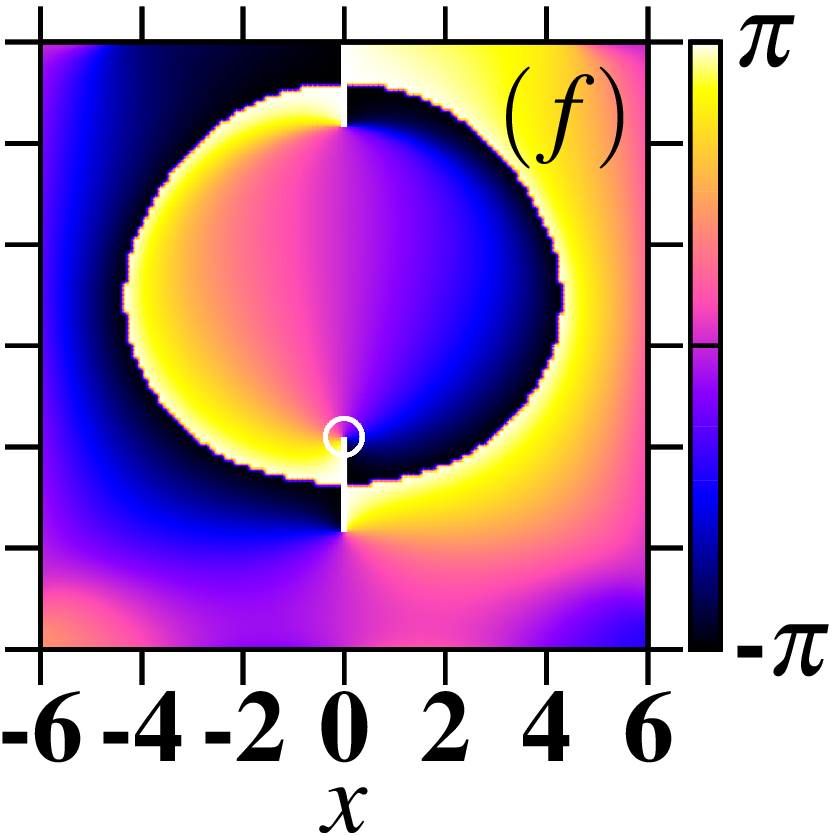}
 \end{tabular}
\caption{The 2D contour plots of densities of components in $z=0$ plane, corresponding to isodensity
contours shown in \ref{Fig6}, for (a) $m_f = +1$,
(b) $m_f = 0$, (c) $m_f = -1$ components of an  asymmetric  soliton with
$c_0 = -10$, $c_1 = -1$ and $\gamma =1$.  The
corresponding phases are shown in (d) for $m_f = +1$, (e) for
$m_f = 0$, and (f) for $m_f = -1$ components.}
\label{Fig6p}
\end{center}
\end{figure}

\begin{figure}[t]
\begin{center}
\includegraphics[trim = 0cm 0cm 0cm 0cm, clip,width=0.45\linewidth,clip]{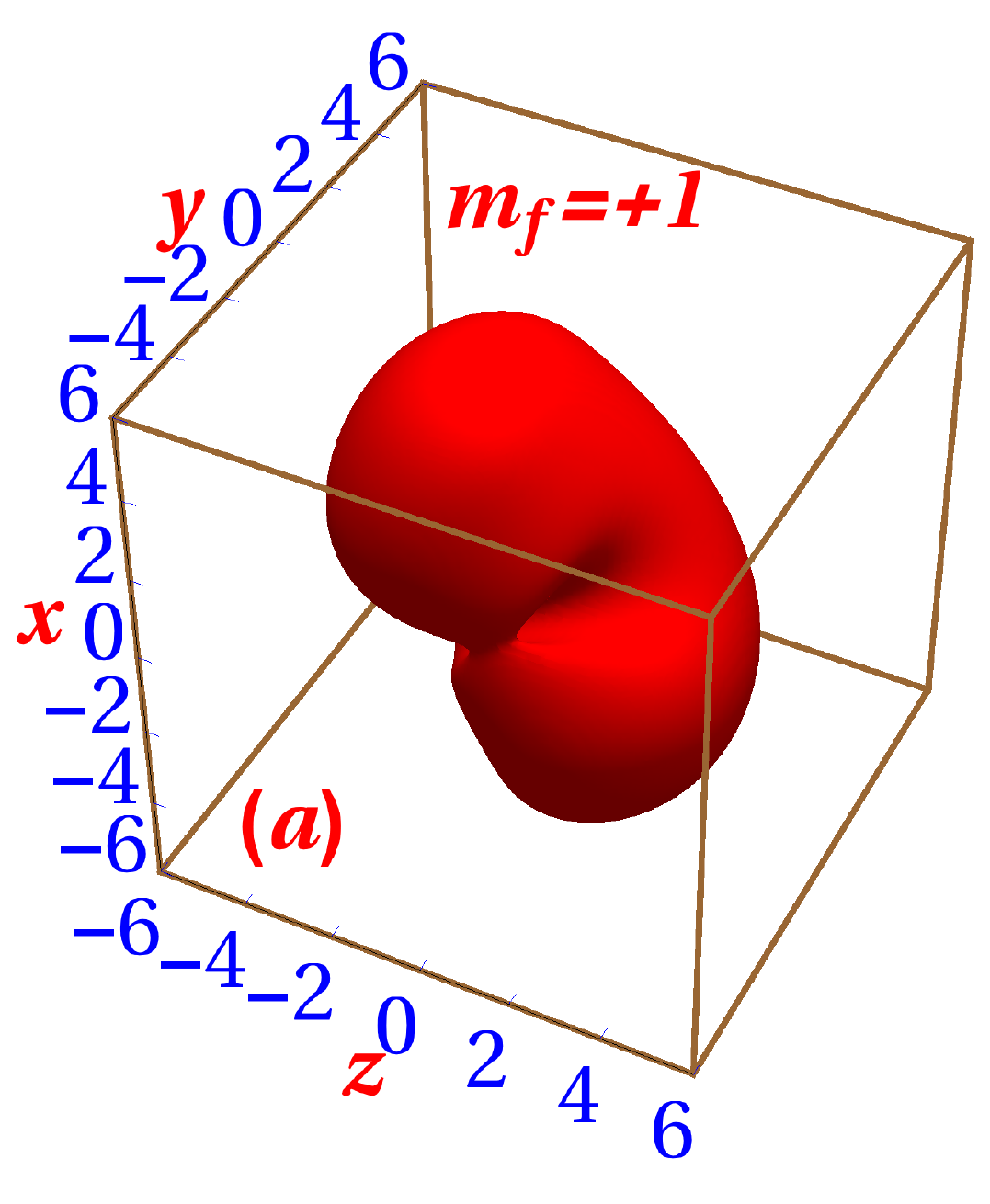}
\includegraphics[trim = 0cm 0cm 0cm 0cm, clip,width=0.45\linewidth,clip]{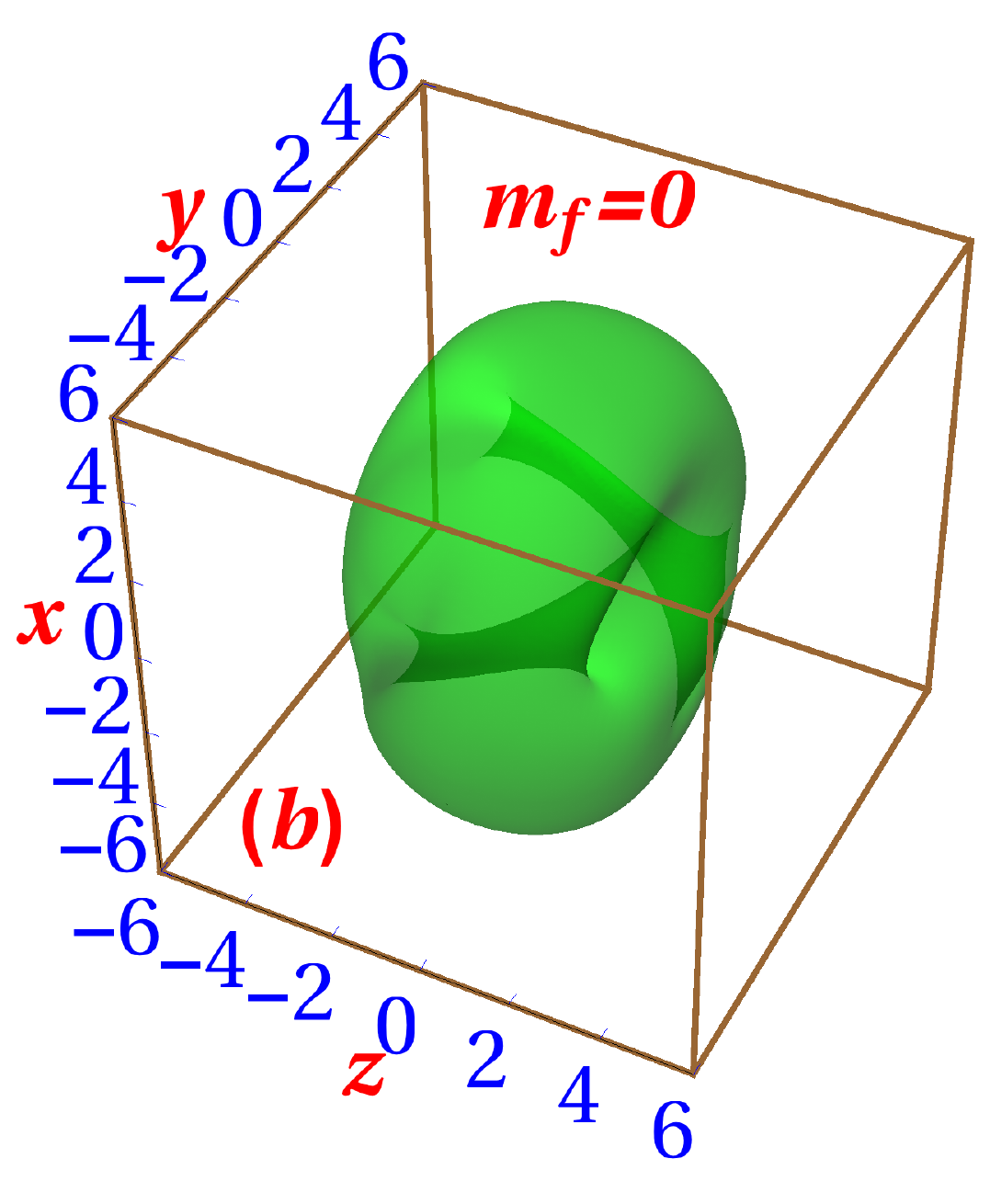}
\includegraphics[trim = 0cm 0cm 0cm 0cm, clip,width=0.45\linewidth,clip]{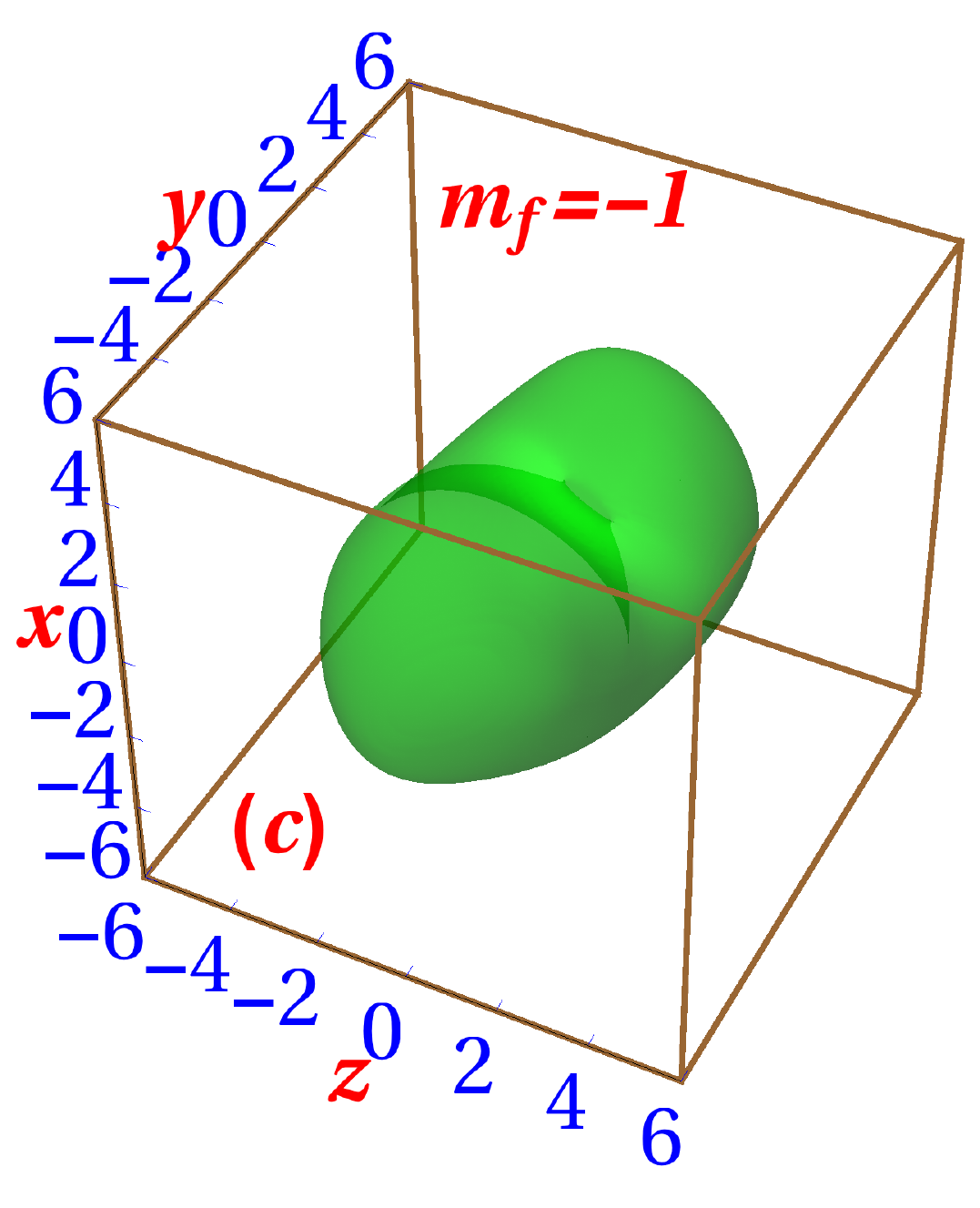}
\includegraphics[trim = 0cm 0cm 0cm 0cm, clip,width=0.45\linewidth,clip]{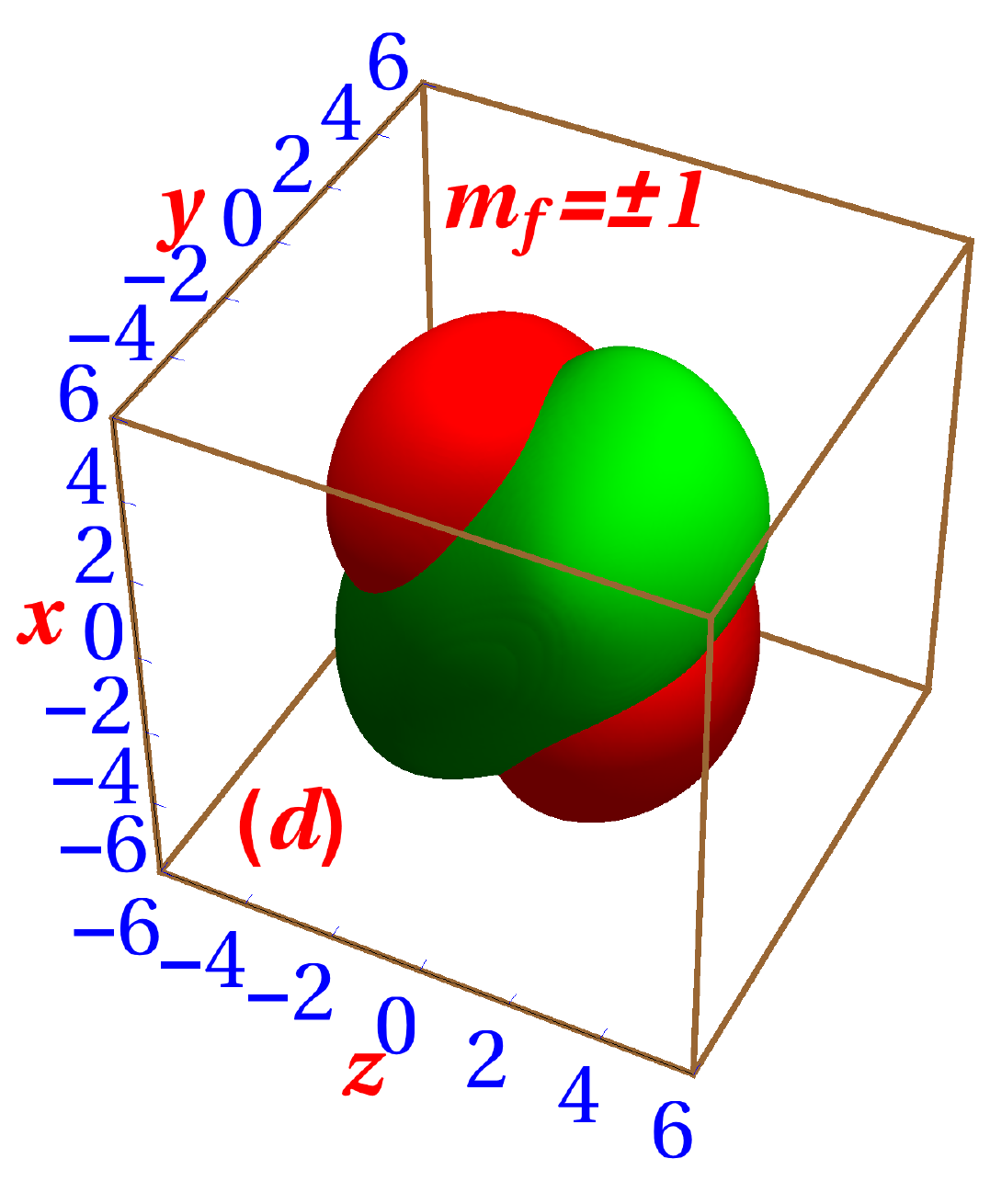}
\caption{Numerical isodensity contour of $|\psi_{m_f}|^2$ for the components 
(a) $m_f = +1$, (b)  $m_f = 0$, (c)  $m_f = -1$ of a moving vortex-bright 
soliton with $c_0 = -10,$ $c_1 = 0.1$, $\gamma =1$, and $v=0.01$ towards 
positive $x$ direction. The density on the contour is $0.00008$. (d) shows 
the isosurfaces in (a) and (c) together. The norms of three components are $0.31$, $0.38$, $0.31$
for $m_f = +1, 0$, and $m_f=-1$, respectively.}
\label{Fig7}
\end{center}
\end{figure}

In the ferromagnetic domain ($c_1<0$) below a critical $c_1$, the system, in 
addition to axisymmetric vortex-bright solitons considered above, can also 
support an asymmetric vortex-bright soliton, which is found to be of lower 
energy than the axisymmetric ones and hence becomes the ground state. The 3D 
numerical isodensity contours of the component wave functions for the 
asymmetric vortex-bright soliton with $c_0 = -10$, $c_1 = -1$ and 
$\gamma =1$ are shown in Figs. \ref{Fig6}(a)-(c). 
 The corresponding 2D contour 
densities and phase profiles in $z=0$ plane are shown in Fig. \ref{Fig6p}(a)-(f). 
 The density on the contour (=0.00008) in Fig. \ref{Fig6} is much smaller than those in Figs. \ref{Fig3}, \ref{Fig4},
and \ref{Fig5} and hence the sizes of solitons in Fig. \ref{Fig6} look larger and these plots  reveal interesting vortex structure in the outer region of low density, which would not have been visible if a larger density on the contour were chosen. 
For  $c_0 = -10$ and 
$\gamma = 1$, the asymmetric solitons exist for $-0.1\le c_1 \le-2.7$.
Below $c_1=-2.7$, no self-trapped solutions exist for $c_0=-10$ and 
$\gamma = 1$, and the system collapses due to an excess of attractive 
interactions. In an asymmetric vortex-bright soliton, we find an antivortex 
line in component $m_f=+1$ (phase winding number $-1$) and a vortex line in 
component $m_f=-1$ (phase winding number $+1$) which are perpendicular to each
other and displaced from the $z$ axis and also from each other. The 
asymmetrically located antivortex and vortex lines in the $m_f=\pm 1$ 
components lead to the kidney-shaped density distributions for these two 
components which fit into each other as shown in Fig. \ref{Fig6}(d). There 
are mutually perpendicular and laterally displaced antivortex and vortex lines
of winding numbers $\mp 1$ in  the $m_f = 0$ component too located in 
regions $y>0$ and $y<0$, respectively, viz. Fig. \ref{Fig6}(b). 
These line 
vortices do not coincide with the line vortices present in the  
$m_f = \pm 1$ components. 
In  Fig. \ref{Fig6p}(d)-(f), the phase-singularities corresponding
to holes (depressions) in 3D isodensity contours of $m_f=0$ $(m_f=\pm 1)$ component are shown enclosed by small
white circles. 
By writing the GP equations 
(\ref{gps-1})-(\ref{gps-2}) in spherical polar coordinates 
($r,\theta, \phi$), it can be seen that these are invariant under the 
transformations: 
$\phi\rightarrow \phi + \delta \phi$ and 
$\psi_{m_f}(r,\theta,\phi)\rightarrow 
\psi_{m_f}(r,\theta,\phi+\delta \phi)e^{-im_f\delta\phi}$;
here $r,\theta,\phi$ are radial, polar and azimuthal coordinates. It implies 
that by rotating the density isosurfaces shown in Fig. \ref{Fig6} about $z$ 
axis, we can get innumerable possible degenerate asymmetric vortex-bright 
solitons. In experiments, this rotation symmetry about $z$ axis will be 
spontaneously broken with the emergence of one of these asymmetric solitons.

\subsection{Dynamically stable moving solitons} 
\label{Sec-IVD} 
\begin{figure}[t]
\begin{center}
\begin{tabular}{lll}
\includegraphics[trim = 0cm 0cm 0cm 0cm, clip,width=0.35\linewidth,clip]{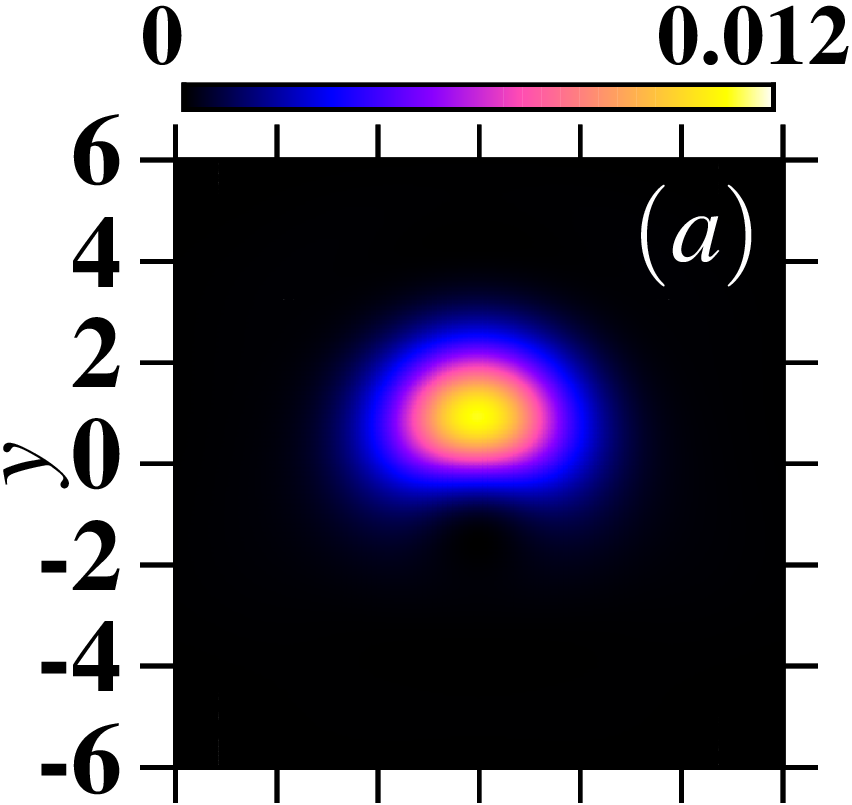}
\includegraphics[trim = 0cm 0cm 0cm 0cm, clip,width=0.28\linewidth,clip]{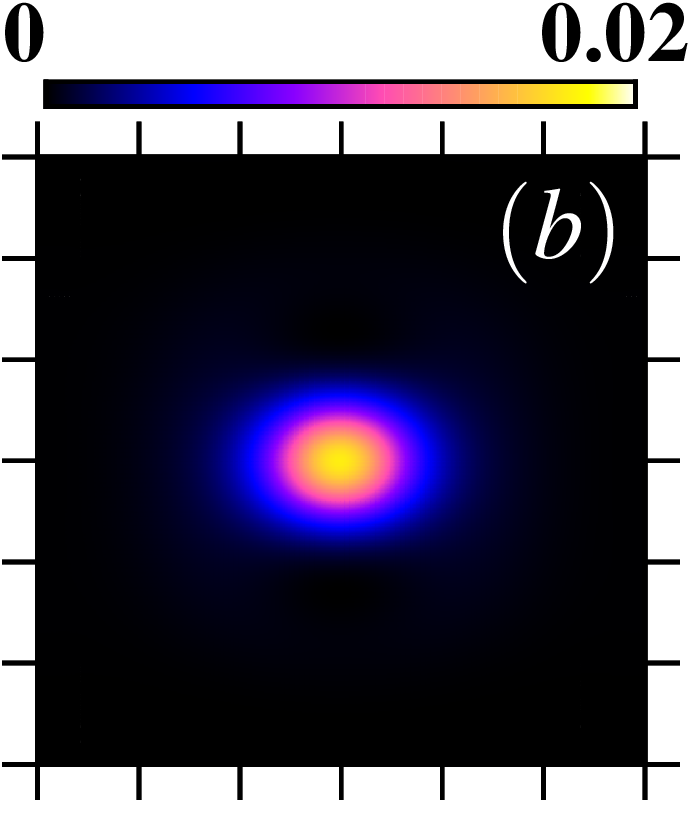}
\includegraphics[trim = 0cm 0cm 0cm 0cm, clip,width=0.295\linewidth,clip]{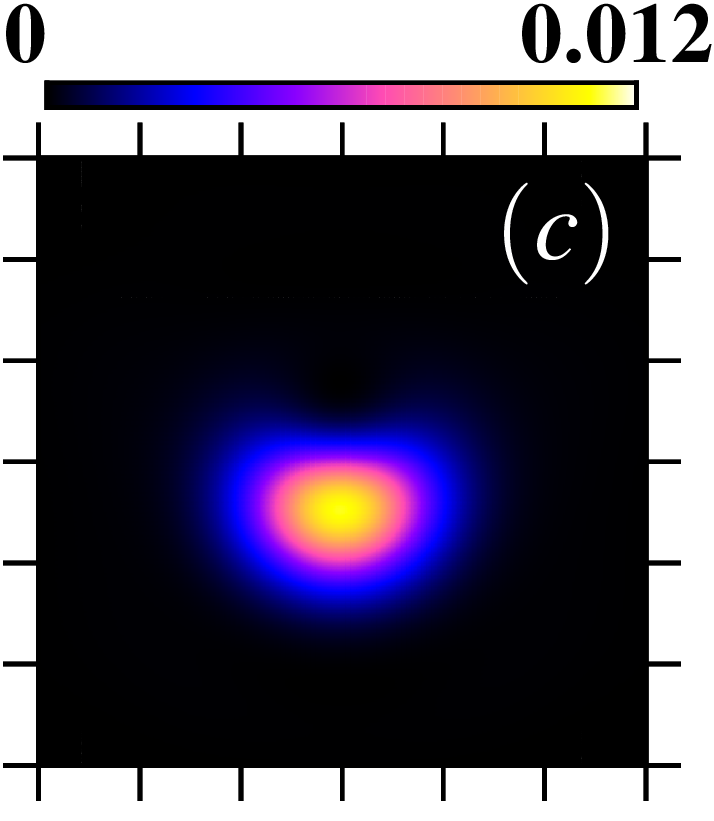}\\
\includegraphics[trim = 0cm 0cm 0cm 0cm, clip,width=0.34\linewidth,clip]{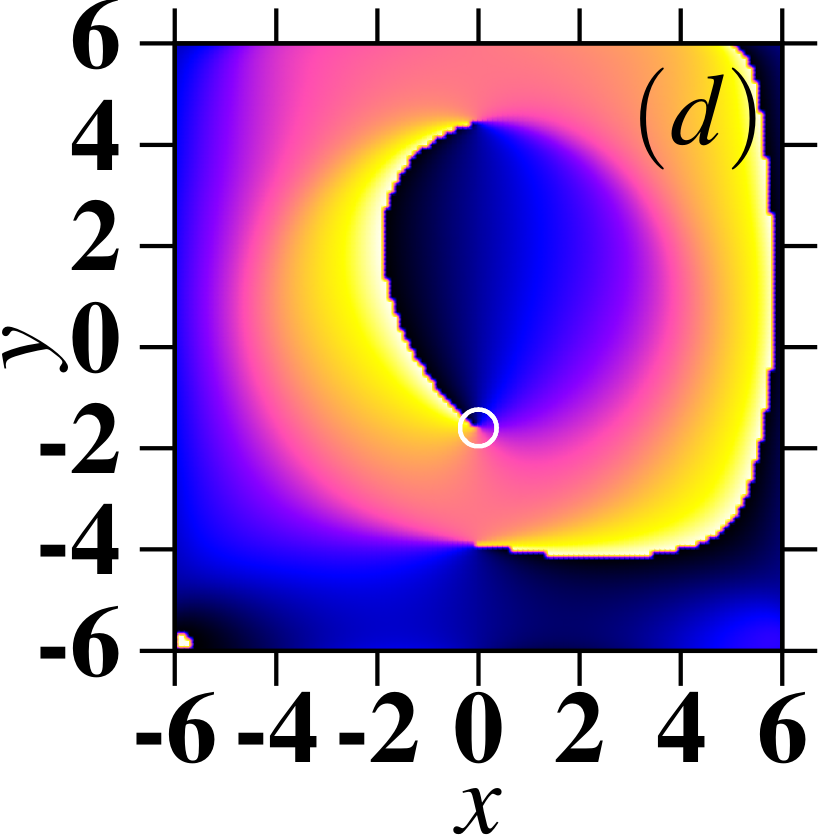}
\includegraphics[trim = 0cm 0cm 0cm 0cm, clip,width=0.285\linewidth,clip]{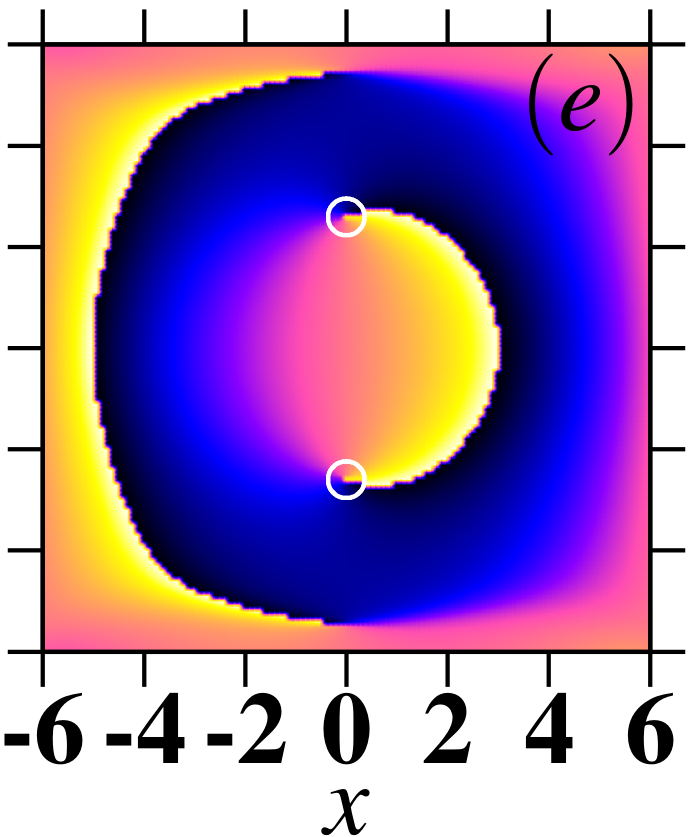}
\includegraphics[trim = 0cm 0cm 0cm 0cm, clip,width=0.345\linewidth,clip]{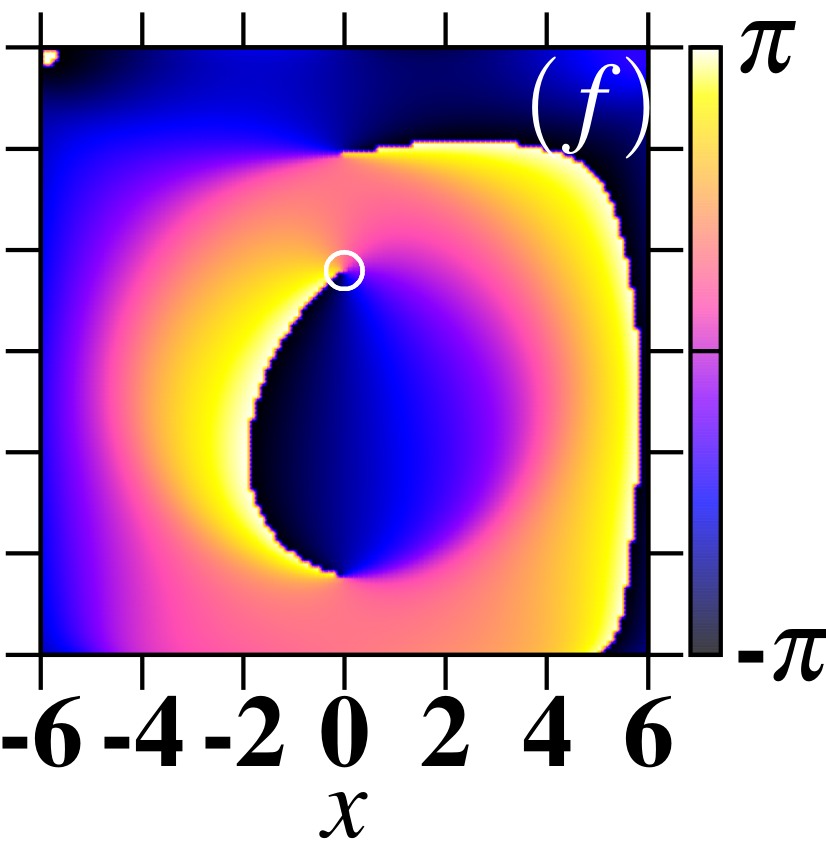}
\end{tabular}
\caption{
The 2D contour plots of densities of components in $z=0$ plane, corresponding to isodensity
contours shown in \ref{Fig7}, for (a) $m_f = +1$,
(b) $m_f = 0$, (c) $m_f = -1$ components of a moving vortex-bright
soliton with $c_0 = -10,$ $c_1 = 0.1$, $\gamma =1$, and $v=0.01$ towards
positive $x$ direction.  The corresponding phases are shown in (d) for $m_f = +1$, (e) for
$m_f = 0$, and (f) for $m_f = -1$ components.
}
\label{Fig7p}
\end{center}
\end{figure}

The GP equations (\ref{gps-1})-(\ref{gps-2}) are not Galilean invariant as
can be shown  by using Galilean transformation 
$x' = x - vt, y' = y, z'=z, t' = t$, where $v$ is the relative velocity along 
$x$ axis of the primed coordinate system with respect to unprimed coordinate 
system, along with the transformation
\begin{equation}
\psi_{j}^{}(x,y,z,t) = \psi'_{j}(x',y',z',t')e^{ivx'+iv^2t'/2},\label{mov-sol}
\end{equation}
in Eqs. (\ref{gps-1})-(\ref{gps-2}). This leads to the following 
Galilean-transformed mean-field GP equations \cite{Gautam-3}
 \begin{align}
i \frac{\partial \psi_{\pm 1}'(\mathbf r')}{\partial t'} &=
 {\cal H}\psi_{\pm 1}'(\mathbf r') 
\pm   c^{}_1F_z'\psi_{\pm 1}'(\mathbf r') 
+ \frac{c^{}_1}{\sqrt{2}} F_{\mp}'\psi_0'(\mathbf r')\label{gpsm-1}\nonumber \\
&-\frac{i\gamma}{\sqrt{2}}\left(
  \frac{\partial\psi_0'}{\partial x'}\mp i\frac{\partial\psi_0'}{\partial y'}
\pm \sqrt{2}\frac{\partial \psi_{\pm 1}'}{\partial z'}\right)
+\frac{\gamma}{\sqrt{2}}v\psi_0',\\
i\frac{\partial \psi_0'(\mathbf r')}{\partial t'} &=
{\cal H}\psi_0'(\mathbf r')  
+ \frac{c_1}{\sqrt 2} [F_{-}'\psi_{-1}'(\mathbf r') 
+F_{+}'\psi_{+1}'(\mathbf r')]\nonumber\\
& -\frac{i\gamma}{\sqrt{2}}\Bigg(\frac{\partial\psi_{1}'}{\partial x} 
+i \frac{\partial\psi_{1}'}{\partial y'} 
 +\frac{\partial\psi_{-1}'}{\partial x'}-i\frac{\partial\psi_{-1}}
{\partial y'}\Bigg)\nonumber \\
& +\frac{\gamma}{\sqrt{2}}v(\psi_{+1}'+\psi_{-1}') 
\label{gpsm-2}.
\end{align}
Due to the $v$-dependent terms in Eqs. (\ref{gpsm-1})-(\ref{gpsm-2}), the 
system is not Galilean invariant. Here for the sake of simplicity, we have 
considered motion along the $x$ axis. In the absence of SO coupling 
($\gamma = 0$), the Galilean invariance is restored, implying that the moving 
solitons, given by Eq. (\ref{mov-sol}), can be trivially obtained by 
multiplying stationary solutions of Eqs. (\ref{gps-1})-(\ref{gps-2}) by 
$e^{i v x}$. This is no longer possible for $\gamma\ne 0$, in which case, the 
moving solitons are the stationary solutions of 
Eqs. (\ref{gpsm-1})-(\ref{gpsm-2}),  presuming that these exist, multiplied 
by $e^{i v x}$ \cite{rela,Sakaguchi,Liu}. As in the case of 
Eqs. (\ref{gps-1})-(\ref{gps-2}), Eqs. (\ref{gpsm-1})-(\ref{gpsm-2}) can be 
solved using imaginary time propagation  with a suitable initial guess for 
component wavefunctions. The 3D isodensity contours of three components for 
a metastable vortex-bright soliton moving with 
velocity $v = 0.01$ along the $x$ axis for $c_0 = -10$, $c_1 = 0.1$ and 
$\gamma = 1$ are shown in Fig. \ref{Fig7}, and the corresponding 
2D contour densities and phase profiles in $z=0$ plane are shown in Fig. \ref{Fig7p}. The
small white circles in the phase profiles in Figs. \ref{Fig7p}(d)-(f) indicate
the phase-singularity responsible for holes or density depressions in 3D isodensity contours 
in Fig. \ref{Fig7}. This moving vortex-bright soliton 
has asymmetric density distribution, whereas the stationary vortex-bright 
soliton for the same set of parameters is a symmetric $(-1,0,+1)$ 
vortex-bright soliton shown in Fig. \ref{Fig3}. The present moving 
vortex-bright soliton has a density distribution very similar to the 
asymmetric soliton of Fig. \ref{Fig6}. The $m_f = +1$ components of both have 
an antivortex of winding number $-1$ and $m_f = -1$ component a vortex of  
winding number $+1$.  
The $m_f=0$ in each has a vortex and an antivortex line
separated from each other.

Notwithstanding these qualitative similarities, there is a crucial difference.
The moving vortex-bright soliton does not have the innumerable degenerate 
counterparts like the asymmetric vortex-bright soliton. This is due the fact 
that the system is no longer invariant under transformations: 
$\phi\rightarrow \phi+\delta \phi$ and $\psi_{m_f}(r,\theta,\phi)\rightarrow 
\psi_{m_f}(r,\theta,\phi+\delta \phi)e^{-im_f\delta \phi}$ due to the presence
of $v$-dependent terms in Eqs. (\ref{gpsm-1})-(\ref{gpsm-2}). However, the 
system is  invariant under transformations:
$\phi\rightarrow \phi+\pi$ and$\psi_{m_f}(r,\theta,\phi)\rightarrow 
\psi_{m_f}(r,\theta,\phi+\pi)e^{-im_f\pi}, v\rightarrow -v$. 
This transformation basically transforms a right moving vortex-bright soliton 
to its degenerate left moving counterpart. For $c_0 = -10,$ $c_1 = 0.1$, and 
$\gamma =1$, the self-trapped solutions of Eqs. (\ref{gpsm-1})-(\ref{gpsm-2}) 
exist for $v\le0.04$.

\section{Summary}

We have studied the formation and dynamics of 3D vortex-bright solitons in a 
three-component SO-coupled spin-1 spinor condensate using variational 
approximation  and numerical solution  of the mean-field GP equation.
The solitons are metastable for $a_0+2a_2<0$ (predominantly attractive) and 
could be stable for $a_0+2a_2>0$ (predominantly repulsive). The ground state 
vortex-bright solitons are axisymmetric of type $(- 1,0, + 1)$ in the polar 
domain, whereas they are fully asymmetric in the ferromagnetic domain below a 
critical strength of spin-exchange interaction. In the latter case, the 
axisymmetric $(- 1,0, +1)$ vortex-bright solitons are the excited states.  
The asymmetric vortex-bright solitons cannot appear in the polar domain. 
In addition, one can have $(0,+1,+2)\equiv(-2,-1,0)$  vortex-bright solitons 
as excited states in both domains. We demonstrate the  dynamical stability of the 
solitons numerically. The present mean-field model is not Galelian invariant, 
and we study moving vortex-bright solitons using a Galilean-transformed 
model. The solitons can move without deformation. 

\label{Sec-V}


\begin{acknowledgements}
S.K.A acknowledges the suport by the Funda\c c\~ao de Amparo \`a Pesquisa do Estado de 
S\~ao Paulo (Brazil) under project 2012/00451-0 and also by 
the Conselho Nacional de Desenvolvimento Cient\'ifico e Tecnol\'ogico (Brazil) under project 303280/2014-0.
\end{acknowledgements}

\end{document}